\documentclass[aps,a4paper,twocolumn,nofootinbib,floatfix,showpacs,preprintnumbers]{revtex4}

\usepackage{amsmath}
\usepackage{amssymb}
\usepackage{epsf}
\newcommand{\beq}{\begin{equation}}
\newcommand{\eeq}{\end{equation}}
\newcommand{\bea}{\begin{eqnarray}}
\newcommand{\eea}{\end{eqnarray}}
\newcommand{\dif}{\mathrm{d}}
\newcommand{\inttau}{\int_0^\beta\!\!\dif\tau\!}
\newcommand{\intV}{\int\!\!\dif^3\!x\;}

\newcommand{\intSTT}{\inttau\intV}
\newcommand{\fslash}[1]{\hbox{$#1$}\!\!\!\!/\;}

\newcommand{\ms}{\overline{\mathrm{MS}}}

\newcommand{\mubar}{\bar{\mu}}

\newcommand{\lih}[1]{\mbox{\boldmath $#1$}}

\newcommand{\sumint}[1]{\hbox{$\sum$}\!\!\!\!\!\!\!\int_{#1}}

\newcommand{\tr}{\mathrm{Tr}}

\begin{document}


\title{Electroweak phase diagram at finite lepton number density}
\author{A. Gynther}
\email[e-mail: ]{antti.gynther@helsinki.fi}
\affiliation{Department of Physical Sciences, Theoretical Physics Division, P.O. Box 64, 00014 University of Helsinki, Finland }

\begin{abstract}
We study the thermodynamics of the electroweak theory at a finite lepton number density. The phase diagram of the theory
is calculated by relating the full $4$-dimensional theory to a $3$-dimensional effective theory which has been previously solved using nonperturbative 
methods. It is seen that the critical temperature increases and the value of the Higgs boson mass at which the first order phase 
transition line ends decreases with increasing leptonic chemical potential.
\end{abstract}

\preprint{hep-ph/0303019}
\preprint{HIP-2003-12/TH}

\maketitle

\section{Introduction}

The complete thermodynamic description of the electroweak theory\footnote{To fix the terminology, in this paper electroweak theory means the electroweak sector of the minimal standard model with
all known physical parameters, essentially $G_\mu,\,m_W,\,m_Z$ and $m_\mathrm{top}$, but parametrized with $m_H$.} depends on only five intensive variables; 
the temperature $T$ of the system, the strength of the external $\mathrm{U}(1)$ magnetic field $\lih{H}_Y$ and the leptonic 
chemical potentials $\mu_{L_i}$. The most studied case is that when only the temperature $T$ and the
conjugated variable entropy are nonzero as it was early understood that at high temperatures the symmetry of the electroweak theory
would be restored \cite{Kirzhnits}. Much work was devoted
to this problem using, for example, perturbative 1-loop \cite{Anderson,Carrington,Dine} and 2-loop
\cite{ArnoldEspinosa} effective potential calculations.
These only work for small scalar self-couplings or for small Higgs
masses and the full solution of the problem required
first a perturbative matching of the full 4-dimensional theory
to an effective 3-dimensional theory \cite{Generic}. The phase
diagram of the effective theory was then numerically solved with lattice Monte
Carlo techniques \cite{ewphdg} (the phase diagram has been studied numerically also with the 4-dimensional theory, see \cite{Csikor:1998eu}). 
The result is that the phase diagram contains a first-order line which ends in a 2nd order critical
point of Ising universality class \cite{univclass}.
Similar techniques were then applied to solve the phase
diagram when also $H_Y$ and the conjugate extensive variable
$VB_Y$ were nonzero \cite{magnetic}. The purpose of this paper is to study the remaining case; how the phase diagram depends on finite
chemical potentials related to lepton and baryon numbers and on the conjugate
extensive variables, net lepton and baryon number densities.

Thermodynamical properties of the electroweak theory at nonzero lepton number density are interesting from many points of
view. Theoretically, the minimal standard model describes Nature to very high accuracy and thus it is important that we know
the theory completely. Especially the partition function is a fundamental concept and to know it under most general circumstances
is of interest. In cosmology, the neutrino degeneracy (the net neutrino number) of the universe is a poorly known number. Best 
limits are given by constraints from big bang nucleosynthesis and cosmic microwave background radiation which limits the degeneracy parameter 
$\xi_\nu \equiv \mu_\nu/T_\nu$ to $\xi_{\mu,\tau}\leq 2.1$ for the muon and tau neutrinos and to $\xi_e \leq 0.3$ for the electron neutrino 
where $\mu_\nu$ are the neutrino chemical potentials and $T_\nu$ is the temperature of the 
neutrino background \cite{Orito}. If such large chemical potentials were present in the very early universe then it raises a question 
about how they affect the electroweak thermodynamics and especially the electroweak phase transition. It has, for example, been proposed that the 
presence of a large lepton number asymmetry might explain the absence of topological defects \cite{topdef} as well as the observed baryon number asymmetry
\cite{baulau}. Finally, a comparison between QCD 
thermodynamics and electroweak thermodynamics is interesting. QCD thermodynamics has of course attracted a lot of interest during the last years
due to experiments carried out at the moment at RHIC in Brookhaven and in the future at LHC in CERN. It should be interesting to see
how the properties of the QCD phase transition as a function of baryonic chemical potential and number of light flavors 
(strange quark mass) compare to the properties of the electroweak phase transition as a function of leptonic chemical potentials and 
number of light bosonic degrees of freedom (Higgs mass).

The role of finite lepton number density in the thermodynamics of the electroweak theory has been discussed already in the literature 
\cite{Linde,Lindecond,BailinLove,Ferrer,Kapusta,Khlebnikov,Laine}. Those studies rely on
perturbative one-loop calculations of the effective potential and the conclusion made is that the critical temperature increases with increasing 
chemical potentials. This can be understood in terms of Bose-Einstein
condensation of the Higgs field due to finite chemical potentials related to gauge charges. The fate of the $W^\pm$ boson condensate,
predicted in \cite{Lindecond}, at high temperatures is also discussed on the same footing \cite{Ferrer,Kapusta}. Vector boson condensation is also discussed in
\cite{Sannino:2002wp}

Purely perturbative calculations are, however, doomed due to infrared divergences and
a nonperturbative study is needed, in general. However, direct Monte Carlo studies of the
full electroweak theory at high temperatures and finite chemical potentials are very difficult for numerous reasons. For example, the system is characterized by
a multitude of scales extending from $\pi T$ (mass scale of nonzero Matsubara modes) to $g^2T$ (mass scale of the magnetic sector of the system). This leads to a need
of large lattices in solving the properties of the system. Furthermore, chiral fermions are notoriously very hard to implement on a lattice. At finite density
there is also the famous ``sign'' problem: the fermionic determinant is complex and thus the integration measure is not positive definite which spoils importance sampling.

We approach this problem by generalizing the successful methods of \cite{Generic,ewphdg} to finite chemical potentials. That is, we calculate the dimensional
reduction of the full 4-dimensional electroweak theory at high temperatures and finite densities to a 3-dimensional effective field theory. The effective theory
can then be solved using Monte Carlo methods in order to find out the phase diagram. This method is, however, by construction limited to small chemical potentials.

The paper is organized as follows. In section \ref{sec2} we define the theories and give the matching between them. In section \ref{sec3} we give 
the results for the phase diagram and in the final section, section \ref{sec4}, we discuss their meaning. Results for the required sum-integrals are given in the Appendix \ref{app}.
     
\section{Dimensional reduction of the electroweak theory at finite chemical potentials}
\label{sec2}
In this section we describe the construction of high temperature effective field theories at finite chemical potential.

\subsection{The fundamental theory}

The electroweak theory at finite temperatures is defined by the Euclidean action
\bea
S & = & \intSTT{\cal L}\quad \quad \quad \mathrm{with} \nonumber \\
{\cal L}  & = &(D_\mu\Phi)^\dagger D_\mu\Phi - \nu^2\Phi^\dagger \Phi + \lambda(\Phi^\dagger \Phi)^2
+ \frac{1}{4}G_{\mu\nu}^aG_{\mu\nu}^a \nonumber \\ 
& & + \frac{1}{4}F_{\mu\nu}F_{\mu\nu}
+ \bar{\lih{l}}_L\fslash{D}\lih{l}_L + \bar{e}_R\fslash{D}e_R + \bar{\lih{q}}_L\fslash{D}\lih{q}_L \\
& & + \bar{u}_R\fslash{D}u_R + \bar{d}_R\fslash{D}d_R + g_Y\left(\bar{\lih{q}}_L\tilde{\Phi}t_R 
+ \bar{t}_R\tilde{\Phi}^\dagger\lih{q}_L \right). \nonumber
\eea 
Here $D_\mu = \partial_\mu + IigA_\mu^a\tau^a + Yig'B_\mu$ where $I$ and $Y$ are the weak isospin and weak hypercharge
of the corresponding doublet/singlet, $\lih{l}_L$ and $\lih{q}_L$ denote the left handed lepton and quark doublets and $e_R,\;u_R$ and $d_R$ denote 
the right handed leptons, up type quarks and down type quarks, respectively. Also, $G^a_{\mu\nu} = \partial_\mu A_\nu^a - \partial_\nu A_\mu^a 
- g\epsilon^{abc}A_\mu^b A_\nu^c$, $F_{\mu\nu} = \partial_\mu B_\nu - \partial_\nu B_\mu$ and 
$\tilde{\Phi} = i\tau^2\Phi^\ast$. Only the top quark is taken to have a nonzero Yukawa coupling. The convention for the Euclidean gamma matrices is as given in \cite{Generic}. 
The bosonic fields ($\phi$) are periodic in $\tau$ while 
fermionic fields ($\psi$) are anti-periodic. Thus they can be expanded in Fourier series (Matsubara modes)
\begin{eqnarray}
\phi(\tau,\lih{x}) & = & \sum_{n=-\infty}^\infty\phi_n(\lih{x})\mathrm{e}^{i2n\pi T\tau}, \\
\psi(\tau,\lih{x}) & = & \sum_{n=-\infty}^\infty\psi_n(\lih{x})\mathrm{e}^{i(2n+1)\pi T\tau}.
\end{eqnarray}
We will employ the power-counting rules $\mbox{$g^{\prime 2} \sim g^3$}$, $\mbox{$\lambda \sim g_Y^2 \sim g^2$}$. The Lagrangian is CP symmetric. Calculations are
performed in Landau gauge.

As well known, the thermodynamics of any system is described by the partition function defined as trace of the density matrix
\begin{equation}
{\cal Z} = \tr\; \mathrm{e}^{-\beta(H-\mu_k N_k)}.
\end{equation}
Here $N_k$ are all the conserved (global or local) charges of the system and $\mu_k$ are
the corresponding chemical potentials. 
In the electroweak theory, at classical level, the lepton number currents and the baryon number current are conserved independently. However, due to the
triangle anomaly, these currents are not conserved in quantum theory (for a review see, e.g. \cite{RubaShapo})
\begin{equation}
\partial_\mu j_\mu \propto g^2\epsilon_{\alpha\beta\mu\nu}G_{\alpha\beta}^aG_{\mu\nu}^a\quad\mathrm{for\;each\;current}.
\end{equation}
It is thus possible to form only $n_f$ conserved linear combinations of these currents. These are usually defined to be
\beq
X_i = \frac{1}{n_f}B - L_i, \quad i=1\dots n_f
\end{equation}
where $n_f$ is the number of families, $B$ is the baryon number and $L_i$ are the lepton numbers for each family
\bea
B & = & \frac{1}{3}\sum_{f,c}\intV \bar{q}_{c,f}\gamma_0 q_{c,f}, \nonumber \\
L_i & = & \intV\left(\bar{e}_i\gamma_0 e_i + \bar{\nu}_i\gamma_0 a_L\nu_i\right) \\
\mathrm{where} & & a_L = \frac{1}{2}(1-\gamma_5). \nonumber
\eea 
Here $f$ and $c$ stand for flavor and color, respectively, and $q_{c,f}$ are the quark fields. The remaining current $n_fB+\sum_iL_i$ is not conserved.

In addition to these globally conserved charges there are locally conserved charges related to the gauge symmetries
of the theory. Of the four gauge generators we can choose two mutually commuting ones for which it is
possible to assign chemical potentials. One must be the hypercharge and as the other one it is convenient to choose the third component of the isospin. The corresponding currents are
\begin{eqnarray}
j_\mu^Y & = & \frac{1}{2}\sum_{\mathrm{fam.}}\Big(\frac{1}{3}\bar{\lih{q}}_L\gamma_\mu\lih{q}_L
+\frac{4}{3}\bar{u}_R\gamma_\mu u_R - \frac{2}{3}\bar{d}_R\gamma_\mu d_R\nonumber \\
& & - \bar{\lih{l}}_L\gamma_\mu\lih{l}_L
-2\bar{e}_R\gamma_\mu e_R\Big) - \frac{i}{2}\left((D_\mu\Phi)^\dagger\Phi - \Phi^\dagger D_\mu\Phi\right), \nonumber \\
j_\mu^3 & = & \frac{1}{2}\sum_{\mathrm{fam.}}\left(\bar{\lih{q}}_L\gamma_\mu\tau^3\lih{q}_L
+\bar{\lih{l}}_L\gamma_\mu\tau^3\lih{l}_L\right) \nonumber \\ 
& & - \frac{i}{2}\left((D_\mu\Phi)^\dagger\tau^3\Phi 
- \Phi^\dagger\tau^3D_\mu\Phi\right) -\epsilon^{3bc}A^{\nu,b} G_{\mu\nu}^c.
\end{eqnarray}
The sums run over the families. Chemical potentials related to gauge charges cannot, however, be chosen freely. In thermal equilibrium the system
must be neutral with respect to gauge charges. This requirement fixes the values of these chemical potentials, which are then functions of temperature
and chemical potentials related to global charges. This can be seen explicitly below.

Taking all the conserved currents into account, the partition function is given by the path integral \cite{Kapusta}
\begin{eqnarray}
{\cal Z} & = & \tr\,\exp\big(-\beta(H-\mu_i X_i-\mu_Y Q_Y - \mu_{T^3}Q_{T^3})\big) \label{eq:trace1}\\
& = &\int\!\!{\cal D}\varphi\exp\left[-\left(S - \inttau\;\sum_{i=1}^{n_f}\mu_i X_i\right)\right] \label{eq:pathint} \\
& \equiv &\int\!\!{\cal D}\varphi\exp\left[-S + \inttau\;\left(\mu_B B + \sum_{i=1}^{n_f}\mu_{L_i} L_i\right)\right] \nonumber
\end{eqnarray}
where $\varphi$ denotes the set of all the fields. Here we have defined ($\tilde{\varphi}$ excludes $B_0$ and $A_0^3$) 
\begin{eqnarray}
\mu_B & \equiv & \frac{1}{n_f}\sum_{i=1}^{n_f}\mu_i, \quad  \mu_{L_i} \equiv -\mu_i, \quad \quad \quad \mathrm{and} \\
S & = & S\left[\tilde{\varphi},B_0+\frac{i\mu_Y}{g'},A_0^3+\frac{i\mu_{T^3}}{g}\right] \label{eq:action}
\end{eqnarray}
with $S$ containing also the gauge fixing and ghost terms.
We see that the constraint $\mbox{$n_f\mu_B + \sum_i\mu_{L_i}=0$}$ is satisfied. 
We also note from Eqs. (\ref{eq:pathint}) and (\ref{eq:action}) that after the integration over the conjugated momentum fields is done (going from Eq. (\ref{eq:trace1}) 
to Eq. (\ref{eq:pathint})), the chemical potentials related to the 
gauge charges enter the path integral the same way as
the static modes of the temporal components of the corresponding gauge fields. Thus, we can interpret them as acting as 
backgrounds for the gauge fields. Therefore, writing $B_0 \rightarrow B_0 + \langle B_0\rangle,\,A_0^3 \rightarrow A_0^3 + \langle A_0^3\rangle$ and
requiring stationarity of free energy with respect to expectation values $\langle B_0\rangle$ and $\langle A_0^3\rangle$ (condition for thermal equilibrium) is 
equivalent to requiring neutrality with respect 
to corresponding gauge charges:
\begin{equation}
0 = \frac{\partial\ln{\cal Z}}{\partial \langle B_0\rangle} = -ig'\frac{\partial\ln{\cal Z}}{\partial\mu_Y} = -\frac{ig' Q_Y}{T}
\end{equation}
and similarly for $\langle A_0^3\rangle$.
It is then convenient to redefine $B_0$ and $A_0^3$ in Eqs. (\ref{eq:pathint}) and (\ref{eq:action}) as
\begin{equation}
g'B_0 \rightarrow g'B_0 - i\mu_Y,\quad gA_0^3 \rightarrow gA_0^3 - i\mu_{T^3}.
\end{equation}
This way the action in Eq. (\ref{eq:action}) becomes the standard electroweak action for these redefined fields and $\mu_Y$ and $\mu_{T^3}$ are not explicit in the path integral
anymore. The chemical potentials can be recovered
as expectation values for the redefined $B_0$ and $A_0^3$ at equilibrium\footnote{To be precise, this requires that the expectation values of the original $B_0$ and $A_0^3$ vanish
in equilibrium. This certainly is the case when the system is neutral.}  
\begin{equation}
\langle B_0\rangle = \frac{i\mu_Y}{g'},\quad \langle A_0^3\rangle = \frac{i\mu_{T^3}}{g}.
\end{equation}
This explicitly shows that these chemical potentials cannot be chosen freely, but are fixed by requiring that the system be in thermal equilibrium.

\subsection{Dimensional reduction at finite $\mu$}

Due to infrared divergences which arise when integrating over the bosonic zero modes $\phi_0(\lih{x})$ (static modes in $\tau$), the path integral (\ref{eq:pathint}) 
cannot be reliably evaluated within perturbation theory. The reason for this is that these modes are light when the temperature is much larger 
than any other mass scale in the theory and therefore the high temperature expansion parameter $g^2T/E$ is large for them. All the other modes (nonstatic in $\tau$), 
$\phi_{n\neq 0}(\lih{x})$ and $\psi_n(\lih{x})$, are on the other hand always very massive,
$m\sim\pi T$ and can therefore be integrated out perturbatively, as can also the static modes with $|\lih{p}|>gT$. We are then led to a very natural idea of formulating 
a three dimensional effective field theory for the static modes
$\phi_0(\lih{x})$ \cite{Generic,Ginsparg}. This effective theory is defined to be the most general theory for the static modes respecting the required symmetries.
It reproduces the Green's functions for these static modes to a controllable accuracy. 

Finite fermion number density affects dimensional reduction in two ways. First, the renormalization of the
fields and parameters as the heavy modes are integrated out changes when compared to the case $\mu = 0$. Second, the symmetries of the fundamental four dimensional 
theory are reduced which gives rise to terms in the effective theories absent at $\mu=0$. More precisely, the introduction of chemical 
potentials to the theory leads to terms in the path integral which break C but preserve P and T thus making the theory, in addition
to being C and P breaking, also CP and CPT breaking. The effective theories may therefore contain
terms which break CP and CPT and which do not appear at $\mu=0$. Such terms must nevertheless still preserve $3$-dimensional gauge- and rotational invariance as well as T
invariance. 

The first effective theory is obtained after integrating out the nonzero Matsubara modes. The resulting
effective theory is a $3$-dimensional $\mathrm{SU}(2)\times \mathrm{U}(1)$ gauge field theory with a fundamental scalar doublet (Higgs) and
four adjoint scalars corresponding to the temporal components of the gauge fields in the fundamental $4$-dimensional theory.
The dimensionally lowest order CP and CPT violating terms arising from the finite chemical potentials in this theory are (in 3-dimensions $[\varphi] = \mathrm{GeV}^{1/2}$)
\begin{figure}
\begin{center}
\mbox{\epsfxsize=8cm\epsffile{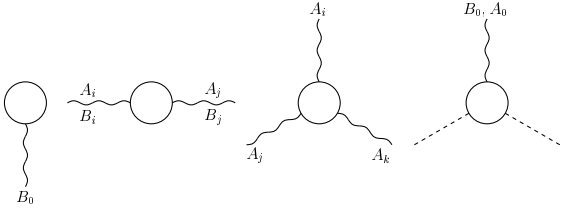}}
\end{center}
\caption{Diagrams leading to new terms in the effective theories. Solid lines correspond to fermions, wavy lines to gauge fields and dashed lines to scalars.}
\label{newterms}
\end{figure}
\bea
\mathrm{dim}\;= & \mathrm{GeV}^{\frac{1}{2}}:& iB_0, \nonumber \\
\mathrm{dim}\;= & \mathrm{GeV}^{\frac{3}{2}}:& iB_0^3,\;\; i\Phi^\dagger A_0^a\tau^a\Phi,\;\; i\Phi^\dagger B_0\Phi, \;\; i B_0 A_0^a A_0^a, \nonumber \\
\mathrm{dim}\;= & \mathrm{GeV}^2:& \epsilon_{ijk}B_i F_{jk}, \nonumber \\
& & \epsilon_{ijk}\left(A_i^a G_{jk}^a - \frac{i}{3}g\epsilon^{abc}A_i^a A_j^b A_k^c\right). \label{eq:nt}
\eea 
which arise from diagrams given in Fig. \ref{newterms}. The factors of $i$ are chosen in such a way that these operators are T invariant (with 
T transformations adopted from the $4$-dimensional theory).
The effective theory does not contain the terms $\tr A_0^a\tau^a$ and $\tr (A_0^a\tau^a)^3$ since these vanish identically due to the
properties of the $\mathrm{SU}(2)$ generators. The coefficient of the term $B_0^3$ is of the order $g'^3\sim g^{9/2}$ and when the matching
of the Green's functions is done to order $g^4$ it can be neglected.

The possibility that there are terms linear in the adjoint scalars in the action of the effective theory is noteworthy. Such terms induce condensates of the corresponding
fields in equilibrium. As already discussed, such condensates for the adjoint scalars are equivalent to nonzero chemical potentials for the gauge charges. 
Thus the emergence of linear terms in the effective theory takes care of neutrality of the system with respect to gauge
charges.

The last two terms of Eq. (\ref{eq:nt}), the so called Chern-Simons terms, are interesting. There of course is a vast literature on the physics induced 
by them (for a review, see \cite{Dunne}). That such terms appear in chiral gauge field theories when fermions are integrated out was first observed 
in \cite{Redlich}. However, in the present study we observe that the coefficients of these terms in the effective theories vanish due to the nonconservation of
$B+L$ ($n_f\mu_B + \sum_i \mu_{L_i}=0$). Thus they do
not play any role in them. The role of those terms at smaller temperatures, where they may be important, has been discussed in \cite{Rubakov}.

The second effective theory is obtained after further integrating out the adjoint scalars, the zero components of the gauge
fields. The resulting theory is a 3-dimensional $\mathrm{SU}(2)\times\mathrm{U}(1)$+Higgs gauge field theory. The form of the theory is fully
determined by the gauge invariance. There cannot, therefore, be any terms in the theory that would be absent at $\mu=0$ as far as symmetries are considered (apart from
the Chern-Simons terms, which nevertheless, as stated above, turn out to be absent). 
Finite chemical potentials show up only in the mapping of the parameters of this theory to the physical variables.

\subsection{Integration over the superheavy modes}

The first effective theory in its most general form is defined by the Lagrangian
\begin{widetext}
\begin{eqnarray}
{\cal L}_1 & = & \frac{1}{4}G_{ij}^a G_{ij}^a + \frac{1}{4}F_{ij}F_{ij} + \left(D_i\Phi\right)^\dagger D_i\Phi 
+ m_3^2\Phi^\dagger\Phi + \lambda_3\left(\Phi^\dagger\Phi\right)^2 
+\frac{1}{2}\left(D_i A_0^a\right)^2  + \frac{1}{2}m_D^2A_0^aA_0^a \nonumber \\
& & + \frac{1}{4}\lambda_A A_0^aA_0^aA_0^bA_0^b + \frac{1}{2}\left(\partial_i B_0\right)^2 + \frac{1}{2}m_D^{\prime 2}B_0^2 
+ h_3 \Phi^\dagger\Phi A_0^aA_0^a + h_3'\Phi^\dagger\Phi B_0^2 + \frac{1}{2}g_3 g_3' B_0 \Phi^\dagger A_0^a\tau^a\Phi \\
& & +\alpha \epsilon_{ijk}\left(A_i^aG_{jk}^a - \frac{i}{3}g_3\epsilon^{abc}A_i^a A_j^b A_k^c\right)  + \alpha'\epsilon_{ijk}B_i F_{jk} + \kappa_1 B_0 + \rho \Phi^\dagger A_0^a\tau^a\Phi 
+ \rho' \Phi^\dagger\Phi B_0 + \rho_GB_0A_0^aA_0^a, \nonumber
\end{eqnarray}
\end{widetext}
where $D_iA_0^a = \partial_iA_0^a-g_3\epsilon^{abc}A_i^bA_0^c$.
The parameters of this theory are to be matched to those of the 4-dimensional theory up to order $g^4$. The factors of $i$ are included in the coefficients $\kappa_1,\;\rho,\;\rho'$ and
$\rho_G$.

In general, theories are related by matching corresponding Green's functions calculated in each theory. If the fields are renormalized
as ($\varphi$ denotes a generic field)
\begin{equation}
\varphi_{3\mathrm{D}}^2 = \frac{1}{T}{\cal Z}_\varphi \varphi_{4\mathrm{D}}^2,
\end{equation}
then the $N$-point Green's functions are related by
\begin{equation}
\Gamma_{3\mathrm{D}}^{\{n_i\}} = \frac{1}{T}\prod_i\left(\frac{T}{{\cal Z}_i}\right)^{n_i/2}\Gamma_{4\mathrm{D}}^{\{n_i\}},
\quad \sum_i n_i = N,
\label{rel}
\end{equation}
where $i$ labels the different fields and $n_i$ is the number of times the field $i$ occurs in the Green's function. This matching 
has been performed for the minimal standard model at zero chemical potentials in \cite{Generic}. At finite fermion number density
those results are modified. 

Let us denote by $\Delta\Gamma_{4\mathrm{D}}$ and $\Delta{\cal Z}_\varphi$ the change in the $4$-dimensional Green's functions and field renormalizations 
due to finite chemical potentials, and by 
$\Delta\Gamma_{3\mathrm{D}}$ the change in the $3$-dimensional Green's functions due to changes in the parameters of the effective theory. We then get
from Eq. (\ref{rel}) that
\begin{eqnarray}
\Delta\Gamma_{3\mathrm{D}}^{(\{n_i\})} & = & T^{N/2-1}\Bigg[\left(\prod_i {\cal Z}_{i,\mu=0}^{-n_i/2}\right)\Delta\Gamma_{4\mathrm{D}}^{(\{n_i\})} \nonumber \\
& & - \left(\sum_i\frac{n_i}{2}\Delta{\cal Z}_i\right)\Gamma_{4\mathrm{D}}^{(\{n_i\})}\Bigg]
\end{eqnarray}
which holds when ${\cal Z}_\varphi$ are calculated up to one-loop order.
This formula allows us to calculate the changes in the mapping of the parameters of the effective theory to physical variables. 

\subsubsection{Changes in the coefficients of the terms present already at $\mu=0$}

\begin{figure}
\mbox{\epsfxsize=8cm\epsffile{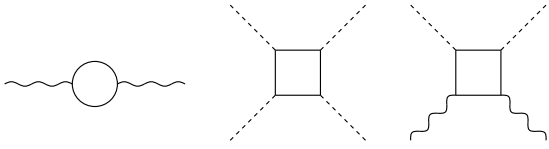}}
\caption{The required diagrams for Debye masses and the couplings. Solid lines correspond to fermions, dashed lines to scalars and wavy lines to gauge bosons.}
\label{coupl}
\end{figure}

\begin{figure}
\mbox{\epsfxsize=8cm\epsffile{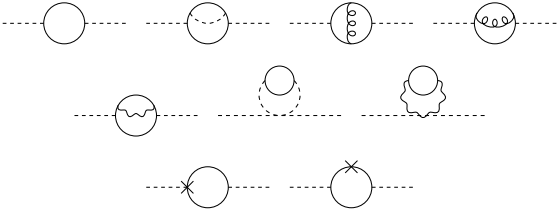}}
\caption{The required scalar 2 point functions for the mass parameter. Solid lines correspond 
to fermions, dashed lines to scalars, wavy lines to electroweak gauge bosons and curly lines to gluons. Crosses denote counterterms.}
\label{sca2pt}
\end{figure}

The field renormalizations can be calculated from the momentum dependent part of the two-point functions. Denoting by ${\cal Z}_\varphi^0$ the
results at zero chemical potential given in Eqs. (141), (142) and (143) of \cite{Generic} we get
\begin{eqnarray}
{\cal Z}_\phi & = & {\cal Z}^0_\phi\left(1-\frac{3g_Y^2}{16\pi^2}{\cal A}\left(\frac{\mu_B}{3}\right)\right), \\
{\cal Z}_{A_0} & = & {\cal Z}^0_{A_0}\left(1-\frac{g^2}{48\pi^2}\left[9{\cal A}\left(\frac{\mu_B}{3}\right) + \sum_{i=1}^{n_f} {\cal A}(\mu_{L_i})\right]
\right), \nonumber \\
{\cal Z}_{A_i} & = & {\cal Z}^0_{A_i}\left(1-\frac{g^2}{48\pi^2}\left[9{\cal A}\left(\frac{\mu_B}{3}\right) + \sum_{i=1}^{n_f} {\cal A}(\mu_{L_i})\right]
\right). \nonumber
\end{eqnarray}
Here we have defined the function ${\cal A}(\mu)$ (see Appendix \ref{app})
\begin{equation}
{\cal A}(\mu) = \psi\left(\frac{1}{2}+\frac{i\mu}{2\pi T}\right) + \psi\left(\frac{1}{2}-\frac{i\mu}{2\pi T}\right) + 2\gamma_E + 2\ln\,4.
\end{equation}
It is now straightforward to find the modifications of the renormalization of the parameters. The Debye masses get a correction coming from
the diagrams in Fig. \ref{coupl}. These are needed only to order $g^2$ (as discussed in \cite{Generic}) and thus the modification of the field renormalizations
do not affect the Debye masses. Denoting by subscript $\mu=0$ the results from Eqs. (160) and (161) of \cite{Generic} we get
\begin{eqnarray}
m_D^2 & = & m_{D,\mu=0}^2 + \frac{g^2}{4\pi^2}\left(\mu_B^2 + \sum_{i=1}^{n_f} \mu_{L_i}^2\right), \nonumber \\
m_D^{\prime 2} & = & m_{D,\mu=0}^{\prime 2} + \frac{{g'}^2}{4\pi^2}\left(\frac{11}{9}\mu_B^2 + 3\sum_{i=1}^{n_f} \mu_{L_i}^2 \right).
\end{eqnarray}
The coupling constants, on the other hand, are modified due to both changes in field renormalizations and changes in loop integrals. The required
diagrams are shown in Fig. \ref{coupl}. The results are
\begin{eqnarray}
\lambda_3 & = & \lambda_{3,\mu=0} -\frac{3g_Y^2T}{16\pi^2}\left(g_Y^2 - 2\lambda\right){\cal A}\left(\frac{\mu_B}{3}\right), \nonumber \\
h_3 & = & h_{3,\mu=0} + \frac{g^4T}{192\pi^2}\left(9{\cal A}\left(\frac{\mu_B}{3}\right) + \sum_{i=1}^{n_f} {\cal A}\left(\mu_{L_i}\right)\right), \nonumber \\
g_3^2 & = & g_{3,\mu=0}^2 +\frac{g^4T}{48\pi^2}\left(9{\cal A}\left(\frac{\mu_B}{3}\right) + \sum_{i=1}^{n_f} {\cal A}\left(\mu_{L_i}\right)\right),
\end{eqnarray}
where $\lambda_{3,\mu=0},\,h_{3,\mu=0}$ and $g^2_{3,\mu=0}$ are given in Eqs. (150), (147) and (146) of \cite{Generic}, respectively.
One may note that $\lambda_A$, the self coupling of the adjoint scalars, does not get any corrections from the chemical potentials and thus
Eq. (162) of \cite{Generic} holds. Also, $h_3' = g_3^{\prime 2}/4$.

Last, the scalar mass parameter is calculated to two loops. One must carefully obtain all the
contributions from fields renormalizations and loop integrals. The diagrams needing recalculation are shown in Fig. \ref{sca2pt}. 
The result is $m_3^2(\mubar) = m_{3,\mu=0}^2(\mubar) + \Delta m_3^2$ where $m_{3,\mu=0}^2(\mubar)$ is given by Eq. (156) of \cite{Generic} and
\begin{widetext}
\begin{eqnarray}
\Delta m_3^2 & = & \frac{g_Y^2(\mubar)}{12\pi^2}\mu_B^2 - \frac{3g_Y^2}{16\pi^2}\left(\nu^2 - \frac{\lambda T^2}{2} - \frac{3g^2T^2}{16} 
- \frac{g_Y^2T^2}{4}\right){\cal A}\left(\frac{\mu_B}{3}\right)
+ \frac{g_Y^2\mu_B^2}{64\pi^4}\left[\frac{3}{4}g^2L_b(\mubar)
-g_Y^2\left(L_f(\mubar)-{\cal A}\left(\frac{\mu_B}{3}\right)\right)\right] \nonumber \\
& & -\Bigg[\left(9g_Y^4+\frac{9}{2}g_Y^2g^2+16g_Y^2g_s^2\right){\cal A}\left(\frac{\mu_B}{3}\right)
-\left(9g_Y^4+\frac{9}{4}g_Y^2g^2-18\lambda g_Y^2
-16g_Y^2g_s^2-\frac{27}{4}g^4\right)16{\cal B}\left(\frac{\mu_B}{3}\right)\Bigg]\frac{T^2}{128\pi^2} \nonumber \\
& & +\left(9g_Y^4+\frac{9}{4}g_Y^2g^2-18\lambda g_Y^2-16g_Y^2g_s^2-\frac{27}{4}g^4\right)\frac{i\mu_B T}{48\pi^3}
\ln\left(\frac{\Gamma\left(\frac{1}{2}-\frac{i\mu_B}{6\pi T}\right)}{\Gamma\left(\frac{1}{2}+\frac{i\mu_B}{6\pi T}\right)}\right) \nonumber \\
& &+\Bigg[9g_Y^2g^2 - 6\left(3g_Y^4-8g_Y^2g_s^2\right)L_b(\mubar)+9g_Y^4L_f(\mubar)
+\left(\frac{9}{2}g_Y^2g^2 + 16g_Y^2g_s^2\right)\left(4\ln\,2 - 1\right) \nonumber \\
& & +\left(9g_Y^4+\frac{9}{4}g_Y^2g^2-18\lambda g_Y^2-16g_Y^2g_s^2-\frac{27}{4}g^4\right)4\gamma_E - \left(9g_Y^4+\frac{9}{2}g_Y^2g^2+16g_Y^2g_s^2\right){\cal A}\left(\frac{\mu_B}{3}\right)\Bigg]
\frac{\mu_B^2}{1152\pi^4} \nonumber \\
& & -\frac{3}{4}g^4\sum_{i=1}^{n_f}\left[\frac{T^2}{8\pi^2}{\cal B}\left(\mu_{L_i}\right)+\frac{i\mu_{L_i} T}{16\pi^3}
\ln\left(\frac{\Gamma\left(\frac{1}{2}-\frac{i\mu_{L_i}}{2\pi T}\right)}{\Gamma\left(\frac{1}{2}+\frac{i\mu_{L_i}}{2\pi T}\right)}\right)
+\frac{\mu_{L_i}^2}{32\pi^4}\gamma_E\right].
\end{eqnarray}
\end{widetext}
Here the function ${\cal B}(\mu)$ (see Appendix \ref{app}) is defined by
\begin{eqnarray}
{\cal B}(\mu) & = & \zeta'\left(-1,\frac{1}{2}+\frac{i\mu}{2\pi T}\right) + \zeta'\left(-1,\frac{1}{2}-\frac{i\mu}{2\pi T}\right) \nonumber \\
& & - 2\zeta'\left(-1,\frac{1}{2}\right)
\end{eqnarray}
and the functions $L_b(\mubar)$ and $L_f(\mubar)$ are as defined in \cite{Generic},

\begin{eqnarray}
L_b(\mubar) & = & \ln\frac{\mubar^2}{T^2} - 2\ln 4\pi + 2\gamma_E \nonumber \\
L_f(\mubar) & = & \ln\frac{\mubar^2}{T^2} - 2\ln \pi + 2\gamma_E.
\end{eqnarray}
Here $\mubar$ is the renormalization scale in the $\ms$ scheme. Note that $\Delta m_3^2$ is independent of $\mubar$ when the running of $g_Y^2$ is taken into account.

\subsubsection{New terms}

\begin{figure}[!b]
\mbox{\epsfxsize=8cm\epsffile{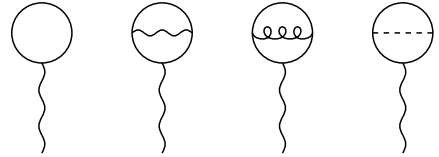}}
\caption{Diagrams contributing to the linear term. Solid lines are fermions, dashed lines scalars, wavy lines electroweak gauge bosons, curly lines are gluons and external legs are $B_0$ legs.}
\label{lintrm}
\end{figure}

As already pointed out, chemical potentials also induce new terms to the effective theories. These arise from the diagrams in Fig \ref{newterms}.
The most important one of these is the term linear in $B_0$ which is related to the neutrality of the system. Calculating to two loops (not including contributions
of the order $g^{\prime 3}\sim g^{9/2}$) we get
\begin{eqnarray}
\kappa_1 & = & -\frac{i\pi}{3}g'T^{5/2}\Bigg[\left(1 - \frac{9g^2}{64\pi^2}\right)\sum_{i=1}^{n_f} \frac{\mu_{L_i}}{\pi T}\left(1+\left(\frac{\mu_{L_i}}{\pi T}\right)^2\right) \nonumber \\
& & -\left(1-\frac{5g_Y^2}{32\pi^2}-\frac{9g^2}{64\pi^2}-\frac{g_s^2}{2\pi^2}\right)\frac{\mu_B}{\pi T}\left(1+\frac{1}{9}\left(\frac{\mu_B}{\pi T}\right)^2\right)\Bigg]. \nonumber \\
\end{eqnarray}
The corresponding diagrams are given in Fig. \ref{lintrm}. The coefficients of the other new terms are needed only to one loop order. The result for them is

\begin{eqnarray}
\rho & = & \frac{i}{8\pi}gg_Y^2T^{3/2}\frac{\mu_B}{\pi T}, \nonumber \\
\rho' & = & -\frac{5i}{24\pi}g'g_Y^2T^{3/2}\frac{\mu_B}{\pi T}, \nonumber \\
\rho_G & = & -\frac{i}{8\pi}g'g^2T^{3/2}\left(\frac{\mu_B}{\pi T}-\sum_{i=1}^{n_f}\frac{\mu_{L_i}}{\pi T}\right), \nonumber \\
\alpha & = & \frac{g^2}{32\pi^2}\left(n_f\mu_B+\sum_{i=1}^{n_f}\mu_{L_i}\right) = 0, \nonumber \\
\alpha' & = & -\frac{g'^2}{32\pi^2}\left(n_f\mu_B+\sum_{i=1}^{n_f}\mu_{L_i}\right) = 0.
\end{eqnarray}
These come from the diagrams in Fig. \ref{newterms}. As already noted, the Chern-Simons terms vanish due to the nonconservation of $B+L$ which sets $n_f\mu_B + \sum_i\mu_{L_i}=0$.

\subsection{One-loop effective potential}

Before integrating over the adjoint scalars $A_0^a$ and $B_0$ it is instructive to consider the effective potential $V_\mathrm{eff}(\varphi,\langle A_0^3\rangle,\langle B_0\rangle)$
for the condensates ($\langle\Phi\rangle = 1/\sqrt{2}\;(0,\;\varphi)^\mathrm{T}$). Although not completely reliable, $V_\mathrm{eff}$ is known to give a rather good description
of the phase transition at small $m_H$ where the scalar self coupling is small. Studies of the electroweak phase transition at finite chemical potentials using perturbatively derived effective
potentials have been carried also previously \cite{Kapusta,BailinLove,Ferrer}.
 
\begin{figure}[!t]
\mbox{\epsfxsize=8cm\epsffile{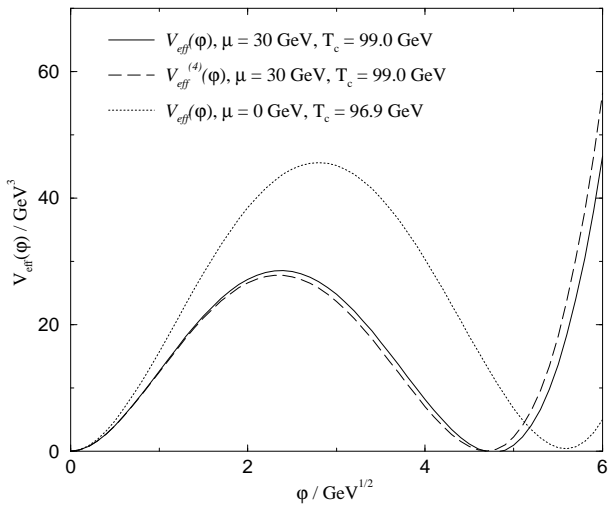}}
\caption{The one-loop effective potentials at the phase transition point for $m_H = 60\, \mathrm{GeV}$. As can be seen, omission of the higher order terms (see Eq. (\ref{eq:defeffpot})) 
has only a small effect.}
\label{oneloopdif}
\end{figure}

At this point we are only interested in the qualitative effects finite chemical potentials can have.
Therefore, in order to simplify the procedure, we neglect the contributions from the terms 
$\lambda_A A_0^aA_0^aA_0^bA_0^b,\;\rho\Phi^\dagger A_0^a\tau^a\Phi,\;\rho'B_0\Phi^\dagger\Phi$ and $\rho_GB_0A_0^aA_0^a$. This is motivated since $\lambda_A\sim g^4$ and 
$\rho_G\sim g^{7/2}$ are small and since the three point vertices $B_0\Phi^\dagger\Phi \sim \rho'$ and $A_0^a \Phi^\dagger\Phi \sim \rho$ are negligible when compared to similar
vertices obtained from the four point vertices $B_0^2\Phi^\dagger\Phi$ and $B_0\Phi^\dagger A_0^a\tau^a\Phi$ after annihilating a $B_0$ leg by the $\kappa_1$ vertex. 
Likewise, we will use $h_3=g_3^2/4,\; h_3'=g_3^{\prime 2}/4$, neglecting higher order corrections, which simplifies some expressions below. This is also consistent with all the
approximations above. Furthermore, we will only consider quantum fluctuations of the magnetic sector $A_i^a,\;B_i$ and of the 
scalars $A_0^1,\;A_0^2$, and treat all the 
condensing scalars $\Phi$, $B_0$ and $A_0^3$ 
only at the tree level. This approximation is adequate to show the effects of chemical potentials. A standard calculation of the one-loop effective potential gives
\begin{widetext}
\begin{eqnarray}
V_\mathrm{eff}(\varphi,\langle B_0\rangle,\langle A_0^3\rangle) & = & \frac{1}{2}m_3^2\varphi^2 + \frac{\lambda_3}{4}\varphi^4 + \frac{1}{2}m_D^2\langle A_0^3\rangle^2 
+ \frac{1}{2}m_D^{\prime 2}\langle B_0\rangle^2 + \frac{1}{2}(h_3\langle A_0^3\rangle^2+h_3'\langle B_0\rangle^2)\varphi^2 
- \frac{1}{4}g_3g_3'\varphi^2\langle B_0\rangle\langle A_0^3\rangle \nonumber \\
& &+ \kappa_1\langle B_0\rangle
-\frac{1}{12\pi}\left(\frac{3}{4}g_3^3(\varphi^2+4\langle A_0^3\rangle^2)^{3/2} + \frac{3}{8}(g_3^2+g_3^{\prime 2})^{3/2}\varphi^3+2(m_D^2 + h_3\varphi^2)^{3/2}\right). \label{eq:oneloop}
\end{eqnarray}
\end{widetext}
Requiring neutrality with respect to gauge charges enforces the conditions
\begin{equation}
\frac{\partial V_\mathrm{eff}}{\partial\langle B_0\rangle} = 0 \quad \frac{\partial V_\mathrm{eff}}{\partial\langle A_0^3\rangle} = 0,
\end{equation}
which gives us
\begin{eqnarray}
\langle B_0\rangle & = & \langle B_0\rangle_0 + \langle B_0\rangle_1, \quad \langle A_0^3\rangle = \langle A_0^3\rangle_0 + \langle A_0^3\rangle_1 \quad \mathrm{with} \nonumber \\
\langle B_0\rangle_0 & = & -\frac{\kappa_1}{m_D^{\prime 2}}\left(1+\frac{h_3}{m_D^2}\varphi^2\right)\frac{1}{1+\left(\frac{h_3}{m_D^2}+\frac{h_3'}{m_D^{\prime 2}}\right)\varphi^2}, \nonumber \\
\langle A_0^3\rangle_0 & = & -\frac{\kappa_1g_3g_3'\varphi^2}{4m_D^2m_D^{\prime 2}}\frac{1}{1+\left(\frac{h_3}{m_D^2}+\frac{h_3'}{m_D^{\prime 2}}\right)\varphi^2}, \nonumber \\ 
\langle B_0\rangle_1 & = & \frac{3}{16\pi}\frac{g_3^4g_3'}{m_D^2m_D^{\prime 2}}\langle A_0^3\rangle_0\varphi^2
\frac{\sqrt{\varphi^2+4\langle A_0^3\rangle_0^2}}{1+\left(\frac{h_3}{m_D^2}+\frac{h_3'}{m_D^{\prime 2}}\right)\varphi^2},\label{eq:conds} \\
\langle A_0^3\rangle_1 & = & \frac{3}{4\pi}\frac{g_3^3}{m_D^2}\langle A_0^3\rangle_0 \left(1+\frac{h_3'}{m_D^{\prime 2}}\varphi^2\right)
\frac{\sqrt{\varphi^2+4\langle A_0^3\rangle_0^2}}{1+\left(\frac{h_3}{m_D^2}+\frac{h_3'}{m_D^{\prime 2}}\right)\varphi^2}. \nonumber
\end{eqnarray}
Here $\langle B_0\rangle_0$ and $\langle A_0^3\rangle_0$ are tree level contributions to the $B_0$ and $A_0^3$ condensates, and $\langle B_0\rangle_1$ and $\langle A_0^3\rangle_1$ are 
corrections to those from the one-loop term of the effective potential. Inserting these back to the effective potential in Eq. (\ref{eq:oneloop}) gives us then the effective potential for
the Higgs expectation value
\begin{eqnarray}
V_\mathrm{eff}(\varphi) & = & \frac{1}{2}\left(m_3^2 -\frac{1}{2\pi}h_3m_D + \frac{h_3'\kappa_1^2}{m_D^{\prime 4}}\right)\varphi^2 \nonumber \\
& & + \frac{1}{4}\left(\lambda_3 -\frac{1}{4\pi}\frac{h_3^2}{m_D} 
- \frac{2h_3'\kappa_1^2}{m_D^{\prime 4}} \left(\frac{h_3}{m_D^2}+\frac{h_3'}{m_D^{\prime 2}}\right)\right)\varphi^4 \nonumber \\
& & - \frac{1}{32\pi}\left(2g_3^3+(g_3^2+g_3^{\prime 2})^{3/2}\right)\varphi^3 + {\cal O}(\varphi^5) \nonumber \\
& \equiv & V_\mathrm{eff}^{(4)}(\varphi) + {\cal O}(\varphi^5). \label{eq:defeffpot}
\end{eqnarray}
The error made by neglecting higher order terms in $\varphi$ is small near the phase transition as can be seen from the example in Fig. \ref{oneloopdif}. 
In terms of the physical parameters we get ($\approx$ corresponds to only tree level matching between couplings of the 4-dimensional theory and physical variables, given for example 
in Eqs. (184)-(185) of \cite{Generic}. The exact one-loop matching, given in Eqs. (183) and (194) of \cite{Generic}, would not change the qualitative considerations.)
\begin{widetext}
\begin{eqnarray}
V_\mathrm{eff}^{(4)}(\varphi) & \approx & \frac{1}{2}\left(-\frac{m_H^2}{2} + \frac{g^2}{16m_W^2}\left(m_H^2+2m_W^2 + m_Z^2 + 2 m_t^2\right)T^2 - \frac{16}{121}\mu^2 
+ T^2{\cal O}\left(\frac{\mu^4}{T^4},g^2\frac{\mu^2}{T^2},g^4\right)\right)\varphi^2 \nonumber \\
& & - \frac{g^3T^{3/2}}{32\pi m_W^3}\left(2m_W^3+m_Z^3\right)\varphi^3 + \frac{T}{4}\left(\lambda + \frac{96}{1331}\frac{\mu^2}{T^2} + {\cal O}\left(\frac{\mu^4}{T^4},g^2\frac{\mu^2}{T^2},g^4\right)
\right)\varphi^4
\end{eqnarray}
\end{widetext}
where we have set all the leptonic chemical potentials equal and $\mu = \mu_{L_i} = -\mu_B$ and only leading order terms in $\mu$ are kept.

\begin{figure}[!t]
\mbox{\epsfxsize=8cm\epsffile{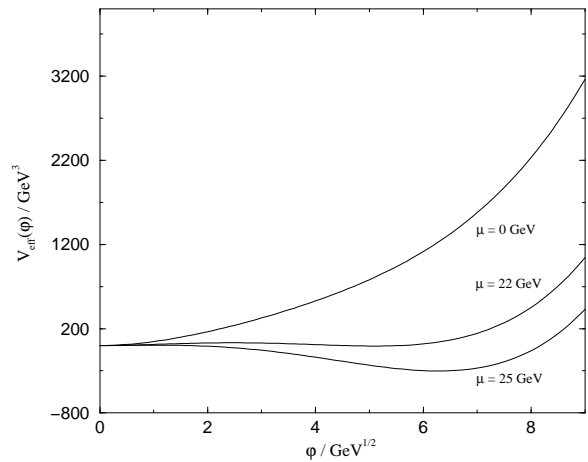}}
\caption{The evolution of the effective potential at fixed temperature $T=98\,\mathrm{GeV}$ and Higgs mass $m_H = 60\,\mathrm{GeV}$ as the chemical potential is increased.}
\label{oneloopdif2}
\end{figure}

This effective potential gives a qualitative picture of the effect of the chemical potentials. First, the critical temperature increases due to decrease of the Higgs mass term. Second,
the scalar self-coupling increases leading to a smaller $\Delta\varphi$ at the transition and thus to a weaker transition. Also the ``barrier'' responsible for the first order phase 
transition is lower.
All these effects can be seen in Fig. \ref{oneloopdif}. In Fig. \ref{oneloopdif2} it can be explicitly seen that finite chemical potentials tend to break the symmetry of the theory.
 
The leading $\mu$-induced corrections to both the scalar mass parameter and self-coupling come from the $B_0$ condensate. Thus, at leading
order the leptonic chemical potentials change the properties of the phase transition through generating nonzero chemical potentials for the gauge charges. These couple to the Higgs field 
and thus change the dynamics of the Higgs field.

We can also note that the $W^\pm$ boson mass is reduced in the broken phase due to the $A_0^3$ condensate 
given in Eq. (\ref{eq:conds}),\footnote{This means that the physics behind the $W^\pm$ condensation is related to the nonzero
chemical potentials for the gauge charges. These chemical potentials couple to the $W^\pm$ bosons and this condensation is nothing but Bose-Einstein condensation due to these chemical potentials.} 
\begin{equation}
m_W^2 \approx \frac{1}{4}\left(g_3^2-\frac{576}{14641}\frac{\mu^2}{T^2}\varphi^2\right)\varphi^2. \label{eq:wmass}
\end{equation}
If $\mu/T$ is large enough, the $W^\pm$ bosons may become unstable leading to a $W^\pm$-condensate. We, however, restrict ourselves to small chemical potentials and study phenomena
near the phase transition where $\varphi^2 \ll T$ and then the $W^\pm$ condensation is not relevant.

Although the perturbatively derived effective potentials serve well in giving a qualitative, and even a quantitative, picture of the phase transition, they cannot be trusted
as the Higgs mass increases. This is clearly seen, for example, in predictions of the nature of the phase transition. Perturbative calculations predict a first order phase transition
for all Higgs masses while nonperturbative studies show that there, in fact, is only a crossover for $m_H\gtrsim 72\;\mathrm{GeV}$ at $\mu=0$. Since the physical Higgs mass is large, it is
important not to rely on perturbative calculations.  

\subsection{Integration over the adjoint scalars}

The phase transition occurs when the scalar mass parameter $m_3^2$ becomes small, $m_3^2 \sim g^4T^2$. Then the adjoint scalars may be considered heavy, $m_D^2 \sim g^2T^2$, and they can 
therefore be integrated out. The resulting effective theory is defined by the Lagrangian
\begin{equation}
{\cal L}_2 = \frac{1}{4}G_{ij}^a G_{ij}^a + \frac{1}{4}F_{ij}F_{ij} + \left(D_i\Phi\right)^\dagger D_i\Phi 
+ \tilde{m}_3^2\Phi^\dagger\Phi + \tilde{\lambda}_3\left(\Phi^\dagger\Phi\right)^2. 
\end{equation}
This reduction step ${\cal L}_1 \rightarrow {\cal L}_2$ is performed in \cite{Generic} in the case $\kappa_1=\rho=\rho'=\rho_G=0$. We now generalize this to finite 
$\kappa_1,\;\rho,\;\rho'$ and $\rho_G$. 
We use as the expansion parameter $h_3/m_D$ and the goal is to calculate the corrections up to order $(h_3/m_D)^2m_D^2$ 
for the scalar mass and to order $(h_3/m_D)^2m_D$ for the couplings in the reduction ${\cal L}_1\rightarrow{\cal L}_2$. To keep track of the contribution of different terms we use the following
power counting rules
\begin{eqnarray}
& & g_3^2 \sim h_3 \sim \frac{h_3}{m_D}m_D,\;\;\; g_3^{\prime 2}\sim h_3' \sim \frac{h_3}{m_D}h_3 \sim\left(\frac{h_3}{m_D}\right)^2m_D, \nonumber \\ 
& & m_D^{\prime 2} \sim \frac{h_3}{m_D}m_D^2,\;\;\; \kappa_1 \sim \left(\frac{h_3}{m_D}\right)^{\alpha-1}m_D^{5/2}, \nonumber \\ 
& &\rho \sim \left(\frac{h_3}{m_D}\right)^{\alpha-1/2}m_D^{3/2},
\;\;\;\rho' \sim \left(\frac{h_3}{m_D}\right)^{\alpha}m_D^{3/2}, \nonumber \\
& &\rho_G \sim \left(\frac{h_3}{m_D}\right)^{2+\alpha}m_D^{3/2}. \label{eq:pc}
\end{eqnarray}
These arise from the power counting rules of the original theory supplemented by setting
$\mu/(\pi T)\sim (h_3/m_D)^\alpha$ for some $\alpha$ where $\mu$ can be any of the chemical potentials. Due to large mass of the top quark we have also relaxed the power counting 
of $\rho$ and $\rho'$ by treating $g_Y^2\sim1$. This protects us from neglecting $\rho$ and $\rho'$ in situations where they would be important. 

\begin{figure}[t!]
\mbox{\epsfxsize=8cm\epsffile{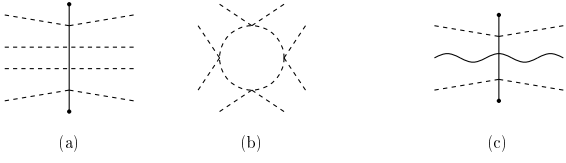}}
\caption{(a) A tree level diagram in the theory given by ${\cal L}_1$ leading to higher order term $(\Phi^\dagger\Phi)^4$ in the last effective theory. 
(b) Leading order contribution to the Green's function
$\langle (\Phi^\dagger\Phi)^4\rangle$ in the theory given by ${\cal L}_2$. (c)  A tree level diagram in the theory given by ${\cal L}_1$ leading to higher order 
term $(\Phi^\dagger\Phi)^2A_i^aA_i^a$ in the last effective theory responsible for the $W^\pm$ condensation . Dashed lines correspond to the fundamental scalar, solid lines to adjoint scalars,  
wavy lines to gauge fields and dots to the $\kappa_1$ vertex.}
\label{treehghord}
\end{figure}

It is essential to keep the chemical potentials sufficiently small in order to keep this
last reduction step meaningful. Consider, for example, the set of diagrams in Fig. \ref{treehghord}a. Such diagrams would lead to terms of the form $(\Phi^\dagger\Phi)^n$, $n\geq 3$ in the final
effective theory. If such higher order terms are not to be included in the effective theory, then the contribution of those terms to the corresponding Green's functions must be sufficiently suppressed
so that they can be neglected. More specifically, in the theory described by ${\cal L}_2$, the leading contribution to the Green's function $\langle(\Phi^\dagger\Phi)^n\rangle$ comes from
the scalar loop in Fig. \ref{treehghord}b and is by naive dimensional calculation of the order $\tilde{\lambda}_3^n\tilde{m}_3^{3-2n} \sim (h_3/m_D)^{3-n}m_D^{3-n}$ near the phase transition where 
$\tilde{m}_3^2\sim (h_3/m_D)^2m_D^2$.
The contribution from the graph in Fig. \ref{treehghord}a on the other hand is of the order of $(h_3/m_D)^{2\alpha+n-3}m_D^{3-n}$. Thus, if we require that the effective theory can reproduce the
Green's functions up to order $g^3\sim(h_3/m_D)^3$ and we neglect the higher order operators produced by the graphs in Fig. \ref{treehghord}a, then we must require that
\begin{equation}
\alpha \gtrsim \frac{3}{2}\quad\mathrm{or}\;\mathrm{equivalently}\quad\frac{\mu}{\pi T} \lesssim g' \label{eq:pc2}.
\end{equation}
This determines the powercounting rules for $\kappa_1,\,\rho,\;\rho'$ and $\rho_G$ which are used when integrating out the adjoint scalars.

Another set of interesting tree level diagrams are those in Fig. \ref{treehghord}c. They would lead to 
terms of the form $(\Phi^\dagger\Phi)^n A_i^aA_i^a$ in the effective theory ${\cal L}_2$. These are interesting since it is these terms that are responsible for the $W^\pm$ condensation, as can, 
for example, be seen from Eq. (\ref{eq:wmass}). There the leading correction to the $W^\pm$ mass is $\sim \varphi^4$ and thus the term in the effective theory ${\cal L}_2$ that would be 
responsible for this correction would be $\sim (\Phi^\dagger\Phi)^2 A_i^a A_i^a$.
The above determined power counting rules allow us to neglect these terms but this means that the effective theory ${\cal L}_2$ cannot predict $W^\pm$ condensation. As already discussed,
this is not a problem when we study phenomena near the phase transition and at small chemical potentials. 

\begin{figure}[!t]
\begin{center}
\mbox{\epsfxsize=8cm\epsffile{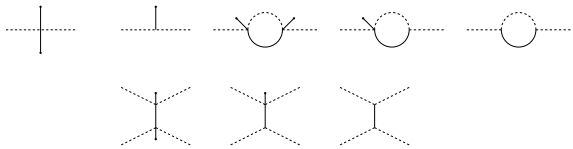}}
\end{center}
\caption{Additional diagrams needed in integration over the adjoint scalars. The dashed lines correspond to the fundamental scalar, solid lines to adjoint scalars and the dot
is the $\kappa_1$ vertex.}
\label{threevert}
\end{figure}

Since the reduction at $\kappa_1=\rho=\rho'=\rho_G=0$ is given in \cite{Generic} we now only need to take into account the contribution from the new terms.
Calculating to the accuracy mentioned before, they only contribute to the scalar mass parameter (up to one-loop level) and scalar self coupling (up to tree level). 
Other contributions are of higher order. The required diagrams are shown in Fig. \ref{threevert}. The results for the parameters are
\begin{widetext}
\begin{eqnarray}
\tilde{m}_3^2 & = & \tilde{m}_{3,0}^2 + \frac{h_3'\kappa_1^2}{m_D^{\prime 4}} - \frac{\rho'\kappa_1}{m_D^{\prime 2}}
- \frac{1}{4\pi}\left[
\left(\frac{3g_3^2g_3^{\prime 2}}{4m_D}+\frac{4h_3^{\prime 2}}{m_D^{\prime}}\right)\frac{\kappa_1^2}{m_D^{\prime 4}} 
- \left(\frac{3g_3g_3'\rho}{m_D}+\frac{4h_3'\rho'}{m_D^{\prime}}\right)\frac{\kappa_1}{m_D^{\prime 2}} + \frac{3\rho^2}{m_D} + \frac{\rho^{\prime 2}}{m_D^{\prime}}\right], \nonumber \\
\tilde{\lambda}_3 & = & \tilde{\lambda}_{3,0} - \frac{1}{2}\left[\left(\frac{g_3^2g_3^{\prime 2}}{4m_D^2}+\frac{4h_3^{\prime 2}}{m_D^{\prime 2}}\right)\frac{\kappa_1^2}{m_D^{\prime 4}} 
- \left(\frac{g_3g_3'\rho}{m_D^2}+\frac{4h_3'\rho'}{m_D^{\prime 2}}\right)\frac{\kappa_1}{m_D^{\prime 2}} + \frac{\rho^2}{m_D^2} + \frac{\rho^{\prime 2}}{m_D^{\prime 2}}\right].
\end{eqnarray}
\end{widetext}
Here $\tilde{m}_{3,0}^2$ and $\tilde{\lambda}_{3,0}$ are as given in Eqs. (174) (first equality) and (169) of \cite{Generic} (where they are denoted by 
$\bar{m}_3^2$ and $\bar{\lambda}_3$). All the other couplings of ${\cal L}_2$ are as given in \cite{Generic} as functions of the parameters of ${\cal L}_1$.

The construction of the theories ${\cal L}_1$ and ${\cal L}_2$ differs in one qualitative aspect. Although neither theory is applicable at high chemical potentials, $\mu > \pi T$, there
nevertheless was no expansion in $\mu/(\pi T)$ in the construction of ${\cal L}_1$. That is, the matching of the parameters of ${\cal L}_1$ to those of the 4-dimensional theory is given to
a certain accuracy in $g^2$ for arbitrary $\mu$. This is not true for the matching between ${\cal L}_1$ and ${\cal L}_2$. There it was essential to assume that the chemical potential is small.
This is easy to understand since some of the couplings of ${\cal L}_1$ are directly proportional to $\mu/(\pi T)$ and thus $\mu$ must be small when ${\cal L}_1$ is studied perturbatively.
We may therefore assume that ${\cal L}_1$ is applicable to somewhat higher chemical potentials than ${\cal L}_2$, up to $\mu \lesssim \pi T$. In the range $g'\pi T \leq \mu \leq \pi T$
the dynamics of ${\cal L}_1$ may be dominated by nonperturbative effects.

\section{The Phase Diagram}
\label{sec3}

The effective theory for the light modes is infrared divergent in perturbation theory and thus it cannot be reliably studied perturbatively.
It has, however, been studied nonperturbatively by Monte Carlo studies in \cite{ewphdg}. 

The theory is parametrized by four parameters. It is convenient to express three of them in a dimensionless form while leaving
one of them to give the energy scale. We define
\beq
x \equiv \frac{\tilde{\lambda}_3}{\tilde{g}_3^2},\quad y \equiv \frac{\tilde{m}_3^2(\tilde{g}_3^2)}{\tilde{g}_3^4}, \quad
z \equiv \frac{\tilde{g}_3^{\prime 2}}{\tilde{g}_3^2}
\end{equation} 
and leave $\tilde{g}_3^2$ to give the dimensions. The value of $z$ is essentially fixed by the Weinberg mixing angle, 
$z \approx \tan^2\theta_W \approx 0.3$. The value of $y$ is tuned to find the phase transition at a fixed value of $x$ which determines the nature
of the phase transition. 

\begin{figure}[t!]
\begin{center}
\mbox{\epsfxsize=8cm\epsffile{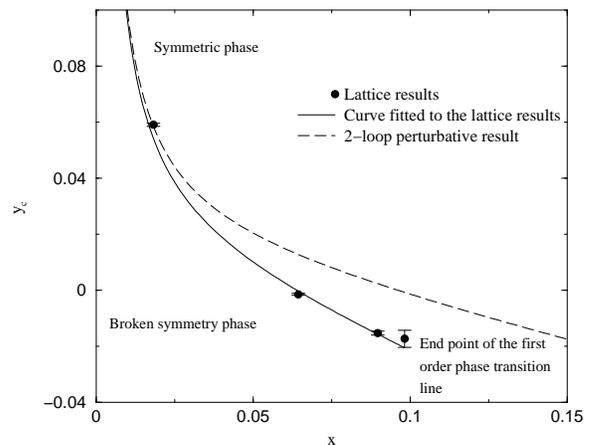}}
\end{center}
\caption{The phase diagram of the effective theory. The results are from \cite{ewphdg} and \cite{univclass}.}
\label{xyphdg}
\end{figure}

The phase diagram of the effective theory is given in Fig. \ref{xyphdg}. The continuous line is a curve fitted to the lattice results which are given
in \cite{ewphdg} and \cite{univclass} for $\mathrm{SU}(2)+$Higgs gauge
theory. The effect of the $\mathrm{U}(1)$ subgroup is to increase the critical $y$ slightly \cite{ewphdg}. The critical line given by a perturbative
calculation is given by the dashed line. As can be seen, the perturbative
result gives quite a good estimate for the value of $y$ at the transition, $y=y_c(x)$, for small $x$. It, however, fails completely in describing the nature of the transition at high $x$. 
Perturbation theory predicts a first order phase transition for all $x$, while Monte Carlo studies have shown that there is a
first order phase transition at small $x$ but that the first order phase transition line has a second order endpoint at 
$x \approx 0.0983$, $y\approx -0.0173$ and for larger $x$ no phase transition is observed \cite{ewphdg,univclass}. Thus, there is
no phase transition for sufficiently large Higgs masses.
    
The phase diagram in Fig. \ref{xyphdg} can be expressed in terms of the physical parameters by the mapping described in section \ref{sec2}. For simplicity 
we set all the leptonic chemical potentials to be equal to each other, $\mu_{L_i} = \mu = -\mu_B$. The theory is
specified by giving the physical parameters the values
$G_\mu=1.664\cdot10^{-5}\, \mathrm{GeV}^{-2},\; m_W=80.42\, \mathrm{GeV},\; m_Z=91.19\, \mathrm{GeV},\; m_t=174.3\, \mathrm{GeV}$ and $\alpha_s(m_Z)=0.118$.
The Higgs mass is left as a free parameter.

\begin{figure}[!t]
\mbox{\epsfxsize=8cm\epsffile{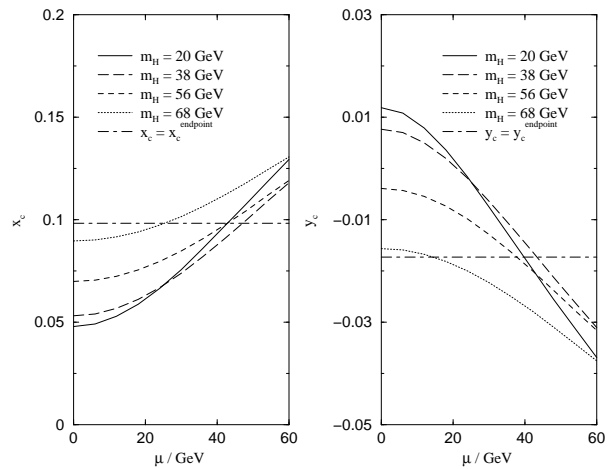}}
\caption{Behavior of $x_c$ and $y_c$ as a function of $\mu$ for different Higgs masses. The regions above the line $x_c=x_c^\mathrm{endpoint}$ and below the line
$y_c=y_c^\mathrm{endpoint}$ correspond to the crossover region of the theory.}
\label{xcyc}
\end{figure}

The exact relation between $(m_H,\; T,\; \mu)$ and $(x,\; y)$ is a complicated function of $\mu$. However, the essential
features of the effect of the chemical potentials can be seen quite easily by just taking the leading corrections due to finite $\mu$
into account. We get
\begin{eqnarray}
x(\mu) & \approx & x(0) + \frac{1}{g^2}\frac{96}{1331}\frac{\mu^2}{T^2} \approx \frac{m_H^2}{8m_W^2} 
+ \frac{1}{g^2}\frac{96}{1331}\frac{\mu^2}{T^2}, \nonumber \\
y(\mu) & \approx & y(0) - \frac{1}{g^4}\frac{16}{121}\frac{\mu^2}{T^2} \nonumber \\
& \approx & -\frac{m_H^2}{2g^4T^2} + 
\frac{1}{g^2}\left(\frac{m_H^2}{16m_W^2}+\frac{3}{16}+\frac{1}{16}\frac{{g'}^2}{g^2}+\frac{1}{4}\frac{g_Y^2}{g^2}\right) \nonumber \\
& & - \frac{1}{g^4}\frac{16}{121}\frac{\mu^2}{T^2}.
\label{xandyatmu}
\end{eqnarray}
Thus the effect of finite $\mu$ is to increase $x$ and decrease $y$. Perturbatively, at tree level the phase transition occurs at $y=y_c=0$. The critical temperature
can then be solved to be
\begin{eqnarray}
T_0^2 & = & \frac{1}{8\lambda + 3g^2 + g'^2 + 4g_Y^2}\left(8m_H^2 + \frac{256}{121}\mu^2\right)
\end{eqnarray}
which explicitly shows that a finite chemical potential increases the critical temperature. At one-loop order the critical line is given by $y_c x = 1/(128\pi^2)$ and expanding this around
the tree level solution $T_c = T_0$ we get for the critical temperature (to first order in $(T_c-T_0)/T_0$)
\begin{equation}
T_c = T_0\left(1 + \frac{1}{16\pi^2}\frac{g^6}{8\lambda + 3g^2 + g'^2 + 4g_Y^2}\frac{1}{\lambda+\frac{96}{1331}\frac{\mu^2}{T_0^2}}\right).
\end{equation}
The factor multiplying the tree level critical temperature decreases as the chemical potentials are increased but that decrease is negligible in comparison
to the simultaneous increase in $T_0$. Thus the critical temperature increases also at one-loop order.

The above reasoning gives a valid qualitative understanding of the behaviour of the system but to obtain quantitatively more reliable results we must map the 
phase diagram in Fig. \ref{xyphdg} to $(T,\,\mu,\,m_H)$ using the complete results from section \ref{sec2}.
Then the values of $x$ and $y$ along the phase transition line, $x_c = x(T_c,\mu,m_H)$ and $y_c=y(T_c,\mu,m_H)$, are given in Fig. \ref{xcyc} as functions of $\mu$. 
It can be observed that as the chemical potentials are increased, the subsequent increase and decrease of $x_c$ and $y_c$, respectively, are fastest at small Higgs masses. This is
easy to understand since $x_c$ and $y_c$ are essentially functions of $\mu/T$. Thus, at small Higgs masses where the critical temperature is lower, increasing $\mu$ leads to a
larger increase in $\mu/T_c$ than at large Higgs masses where the critical temperature is higher. Therefore the changes in $x_c$ and $y_c$ are also larger at smaller Higgs masses.  

\begin{figure}[t!]
\mbox{\epsfxsize=8cm\epsffile{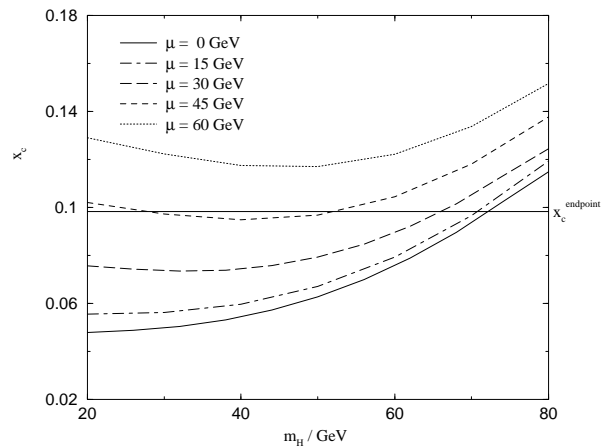}}
\caption{The behavior of $x_c$ as a function of $m_H$.}
\label{xh}
\end{figure}

This has an interesting consequence. The $x_c(\mu)$ curves for different $m_H$ intersect and for sufficiently large chemical potentials the value of $x_c$ is,
in fact, a decreasing function of the Higgs mass (at least for sufficiently small Higgs masses). This can be seen explicitly in Fig. \ref{xh}. Therefore, under these specific conditions, 
the phase transition appears
to become stronger as the Higgs mass is increased (again, at least as long as the Higgs mass remains sufficiently small). This, however, does not seem to have any physical
relevance. Dimensional reduction is not reliable at small Higgs masses where this effect is strongest. As the Higgs mass is increased also the chemical potentials must be increased
in order to recover this anomalous behaviour. However, at these larger Higgs masses and chemical potentials, the value of $x_c$ is above the endpoint value $x_c^\mathrm{endpoint}=0.0983$
and there is therefore no phase transition. Especially, at physical Higgs masses, $m_H \gtrsim 115\,\mathrm{GeV}$, there is no phase transition.

\begin{figure}[!t]
\mbox{\epsfxsize=8cm\epsffile{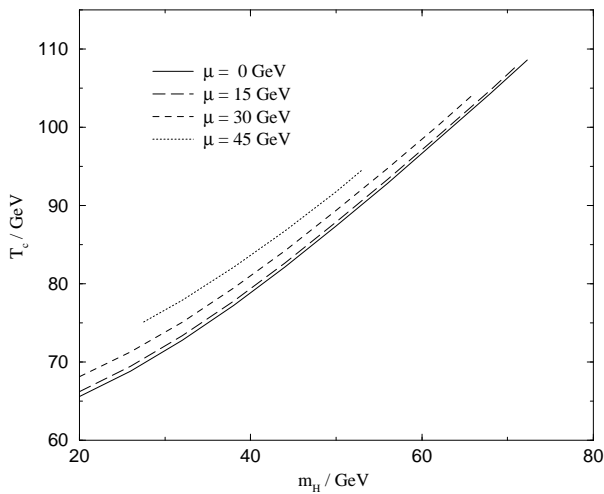}}
\caption{The electroweak phase diagram on the $m_H-T$ plane.}
\label{mhTphdg}
\end{figure}

\begin{figure}[!t]
\mbox{\epsfxsize=8cm\epsffile{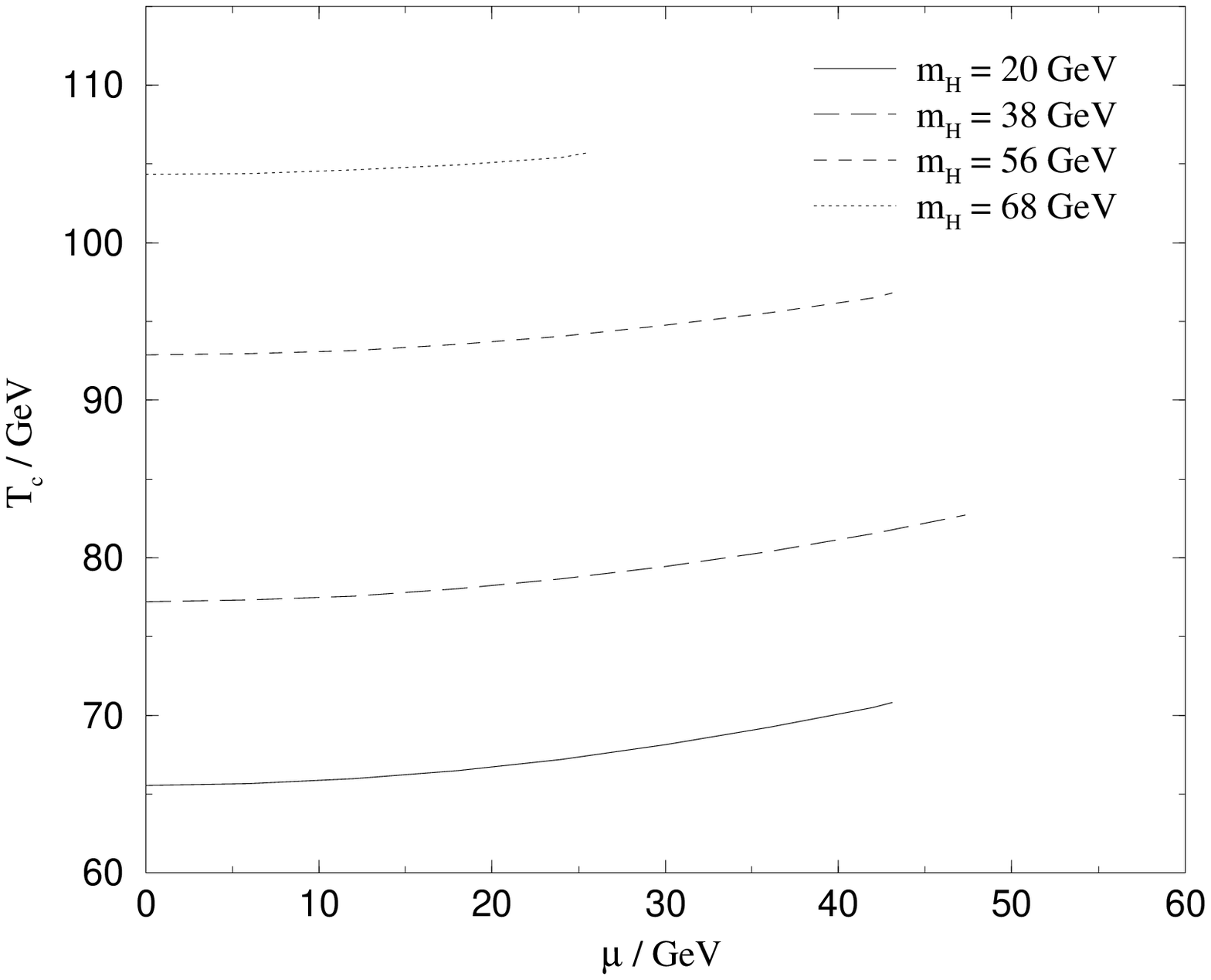}}
\caption{The electroweak phase diagram on the $\mu-T$ plane.}
\label{mhTphdg2}
\end{figure}

The phase diagram in terms of the physical parameters is given in Figs. \ref{mhTphdg} and \ref{mhTphdg2}. The qualitative picture based on perturbation theory can be seen to be correct. The
critical temperature increases with $\mu$. Some interesting thermodynamics can be deduced from this. According to the Clausius-Clapeyron relations
\begin{equation}
\frac{dT}{d\mu} = -\frac{n_s - n_b}{s_s-s_b},
\label{cc}
\end{equation}
where $dT/d\mu$ is measured along the phase transition line, and $n_{s,b}$ and $s_{s,b}$ are the lepton number and entropy densities in each phase, respectively. Since $dT/d\mu$ is
positive, the lepton number difference between the phases is of the opposite sign than the entropy difference. Furthermore, since entropy is higher in the symmetric phase (the
high temperature phase) it means that the lepton number density is higher in the broken symmetry phase.

The behaviour of the endpoint of the first order phase transition line is also of interest. For a fixed Higgs mass, a finite chemical potential strengthens the scalar self-coupling 
at the transition point as seen from Fig. \ref{xcyc}. Thus, increasing $\mu$ weakens the transition and the value of the critical Higgs mass where the first order
phase transition line ends in a second order endpoint is a decreasing function of $\mu$. Especially, there cannot be a first order electroweak phase transition at physical Higgs masses 
in the minimal standard model even when $\mu \neq 0$. Moreover, for sufficiently high chemical potential there appears to be no phase transition for any value of $m_H$.  
The location of the second order endpoint in terms of the Higgs mass and critical temperature are given in table
\ref{eptable} for some values of the chemical potential.

\begin{figure}[!t]
\mbox{\epsfxsize=8cm\epsffile{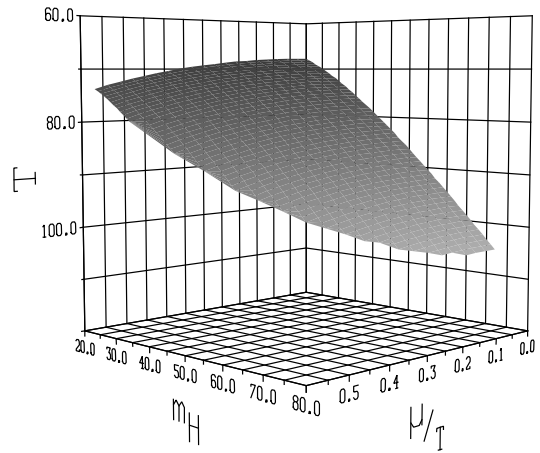}}
\caption{The phase diagram. Note that the temperature grows downwards}
\label{phdg3d2}
\end{figure}

\begin{table}
\centering
\begin{tabular}{|c|c|c|} \hline
$\mu$  & $m_H^\mathrm{end\;point}$ & $T_c^\mathrm{end\;point}$  \\ \hline
$0\;\mathrm{GeV}$ & $72\;\mathrm{GeV}$ & $109\;\mathrm{GeV}$ \\ \hline
$15\;\mathrm{GeV}$ & $71\;\mathrm{GeV}$ & $108\;\mathrm{GeV}$ \\ \hline
$30\;\mathrm{GeV}$ & $66\;\mathrm{GeV}$ & $104\;\mathrm{GeV}$ \\ \hline
$45\;\mathrm{GeV}$ & $52\;\mathrm{GeV}$ & $94\;\mathrm{GeV}$ \\ \hline
\end{tabular}
\caption{The location of the endpoint of the first order phase transition line.}
\label{eptable}
\end{table}

\section{Discussion}
\label{sec4}

In this paper we have determined quantitatively how the equilibrium phase diagram of the electroweak theory, parametrized by the Higgs mass, depends
on the temperature and small leptonic chemical potentials. In particular, we have studied the change of the second order endpoint of the first order phase transition
line. It is seen that the critical temperature increases and the critical value of Higgs mass where the first order phase transition line ends decreases 
as the chemical potentials are increased. These results are summarized in Fig. \ref{phdg3d2}.

It is interesting to qualitatively compare these results for the critical temperature $T=T_c(m_H,\mu)$ of the electroweak phase transition to 
the critical temperature $T=T_c(m_s,\mu_B)$, $m_s = $ strange quark mass, $\mu_B = $ baryonic chemical potential, of the QCD phase transition. In QCD there is also a first order
critical line ending in a 2nd order point: if $\mu_B=0$ then for $m_s=0$ ($N_f=3$) there is a 
first order phase transition while for $m_s=\infty$ ($N_f=2$) there is only a crossover. Thus, the situation on the $m_s-T$ plane is like in Fig. \ref{mhTphdg}. However, in contrast to 
Fig. \ref{mhTphdg2}, on the $\mu_B-T$ plane the
critical lines for different $m_s$ bend downwards starting from $\mu_B=0$ (in Fig. \ref{mhTphdg} the $\mu_B>0$ curves would reside below the $\mu_B=0$ curve). 
As discussed after Eq. (\ref{cc}), this implies that in the QCD phase transition the entropy density and the baryon number density change in a similar way. Also, in QCD, when $m_s$ is
large enough, there is a first order phase transition only if $\mu_B$ is greater than some critical value $\mu_{B,c}(m_s)$, again in contrast
to Fig. \ref{mhTphdg2} which says that there is a first order electroweak phase transition only if $\mu$ is less then some critical value $\mu_c(m_H)$.   

The effective theory used in this paper to study the electroweak phase transition cannot be used at large chemical potentials. Therefore, it cannot, for example, 
predict the emergence (or perhaps lack
of it) of the $W^\pm$ condensate. The reason for this is that the driving force behind the $W^\pm$ condensation, Bose-Einstein condensation, would appear in this
theory through higher order terms of the form $(\Phi^\dagger\Phi)^nA_i^aA_i^a,\;n\geq 2$. Those terms would modify the $W^\pm$ mass in such a way that the condensation
might occur. We have, however, neglected such higher order terms from the theory which is possible if the chemical potentials are small. The $W^\pm$ condensation
is in any case a high $\mu$ phenomenon and thus our approximation is self consistent.

\section*{Acknowledgements}

The author would like to thank K. Kajantie for suggesting the topic and for numerous discussions during the work. Discussions with M. Laine and A. Vuorinen are also gratefully acknowledged.
The work has been supported by the Graduate School for Particle and Nuclear Physics, GRASPANP.

\appendix

\section{Sum-integrals at finite $\mu$}
\label{app}

In this appendix we give the results for the required fermionic sum-integrals for arbitrary $\mu$. Similar integrals also appear in the context 
of computing QCD quark number susceptibilities \cite{aleksi}.

\subsection{One-loop integrals}

At one-loop we have two types of integrals. Both of these are easily evaluated by first doing the $3-2\epsilon$ dimensional momentum integration
and then performing the sum using Riemann zeta functions. Denoting the sum-integrals by
\begin{equation}
\sumint{p} \equiv \left(\frac{\mathrm{e}^{\gamma_E}\mubar^2}{4\pi}\right)^\epsilon\sum_{n=-\infty}^\infty\int\!\!\frac{d^{3-2\epsilon}p}{(2\pi)^{3-2\epsilon}},\quad p_0 = (2n+1)\pi T -i\mu,
\end{equation}
where $\mubar$ is the scale in the $\ms$ scheme, we get
\begin{eqnarray}
I^{2n}_\alpha & \equiv & \sumint{p_f}\frac{p_0^{2n}}{(p^2)^\alpha} \nonumber \\
& = &\frac{\Gamma(\alpha-3/2+\epsilon)}{8\pi^{3/2}\Gamma(\alpha)}\left(\mathrm{e}^{\gamma_E}\mubar^2\right)^\epsilon
T(2\pi T)^{2n+3-2\epsilon-2\alpha} \nonumber \\
& & \times 2\mathrm{Re}\left[\zeta\left(2\alpha-2n-3+2\epsilon,\frac{1}{2}-\frac{i\mu}{2\pi T}\right)\right],  \nonumber \\
I^{2n+1}_\alpha & \equiv & \sumint{p_f}\frac{p_0^{2n+1}}{(p^2)^\alpha} \nonumber \\
 & = & \frac{\Gamma(\alpha-3/2+\epsilon)}{8\pi^{3/2}\Gamma(\alpha)}\left(\mathrm{e}^{\gamma_E}\mubar^2\right)^\epsilon
T(2\pi T)^{2n+4-2\epsilon-2\alpha} \nonumber \\
& & \times 2i\mathrm{Im}\left[\zeta\left(2\alpha-2n-4+2\epsilon,\frac{1}{2}-\frac{i\mu}{2\pi T}\right)\right], \nonumber \\ \label{a1}
\end{eqnarray}
where
\begin{equation}
\zeta(z,q) = \sum_{n=0}^\infty\frac{1}{(n+q)^z},\quad z>1 \label{zetadef}
\end{equation}
is the generalized zeta function. Its properties are discussed below.
Especially the following special cases, expanded around $\epsilon = 0$, are needed:
\begin{widetext}
\begin{eqnarray}
I^0_1 & = & -\frac{T^2}{24}\left\{1 + \frac{3}{\pi^2}\frac{\mu^2}{T^2} + \epsilon\left[
\left(1 + \frac{3}{\pi^2}\frac{\mu^2}{T^2}\right)\left(L_f(\mubar)-2\gamma_E+2-4\ln\;2 \right) \right.\right. \nonumber \\
& & \left. \left.  +24\zeta'\left(-1,\frac{1}{2}-\frac{i\mu}{2\pi T}\right)
+ 24\zeta'\left(-1,\frac{1}{2}+\frac{i\mu}{2\pi T}\right)\right]\right\} + {\cal O}(\epsilon^2), \label{eq:zetas} \\
I^0_2 & = & \frac{1}{16\pi^2}\left[\frac{1}{\epsilon} + L_f(\mubar) - 2\gamma_E-4\ln\;2 - \psi\left(\frac{1}{2}-\frac{i\mu}{2\pi T}
\right)
-\psi\left(\frac{1}{2}+\frac{i\mu}{2\pi T}\right)\right] \nonumber \\
& & + {\cal O}(\epsilon), \label{eq:psis} \\
I^1_1 & = & \frac{i\pi T^3}{12}\frac{\mu}{\pi T}\left(1+\frac{\mu^2}{\pi^2T^2}\right) + {\cal O}(\epsilon), \\
I^1_2 & = & \frac{iT}{8\pi}\frac{\mu}{\pi T} + {\cal O}(\epsilon),
\end{eqnarray}
\end{widetext}
where the primes in zeta functions denote derivatives with respect to the first argument and $\psi(z) \equiv \partial_z\ln\Gamma(z)$.
The results are exact in $\mu$. Terms of the order $\epsilon$ are needed only for the first sum-integral.

\subsection{Two-loop integrals}

Most of the needed two-loop sum-integrals can be reduced to products of one-loop sum-integrals. Only the sum-integral
corresponding to the ``setting sun'' diagram must be evaluated explicitly. In calculating it we follow the procedure outlined in
\cite{ArnoldZhai}. 

The integral we are interested in is
\begin{eqnarray}
I_\mathrm{sunset} & \equiv & \sumint{p_f,k_b}\frac{1}{p^2k^2(p+k)^2} = \sumint{k_b}\frac{1}{k^2}\Pi^f(k) \\
p_0 & = & (2n+1)\pi T-i\mu,\quad k_0 = 2m\pi T,\quad n,m \in \mathbb{Z}, \nonumber
\end{eqnarray}
where we have defined the integral $\Pi^f(k)$ as in \cite{ArnoldZhai}. By going to configuration space where the propagator is given by (at $\epsilon=0$)
\begin{equation}
\Delta(p_0,\lih{r}) = \frac{\mathrm{e}^{-|p_0|r}}{4\pi r},
\end{equation}
this integral becomes
\begin{eqnarray}
\Pi^f(k) & = & T\sum_{\{p_0\}}\int\!\!\dif^3\!r\;\mathrm{e}^{i\lih{k}\cdot\lih{r}}\Delta(p_0,\lih{r})\Delta(p_0+k_0,\lih{r}) \nonumber \\
& = & \frac{T}{16\pi^2}\!\!\sum_{\{p_0\}}\!\int\!\!\dif^3\!r\;\frac{\mathrm{e}^{i\lih{k}\cdot\lih{r}}}{r^2}\mathrm{e}^{-|p_0|r}\mathrm{e}^{-|p_0+k_0|r}.
\end{eqnarray}
Here we have used the notation of \cite{ArnoldZhai} where $\sum_{\{p_0\}}$ means summation over fermionic Matsubara modes.

The influence of the chemical potentials resides now in the sum
\begin{equation}
\sum_{\{p_0\}}\mathrm{e}^{-|p_0|r}\mathrm{e}^{-|p_0+k_0|r} = \mathrm{e}^{-|k_0|r}\left(\frac{\cos(2\mu r)}{\sinh(2\pi T r)} + \frac{|k_0|}{2\pi T}\right)
\end{equation}
and thus we get
\begin{equation}
\Pi^f(k) = \frac{T}{16\pi^2}\int\!\!\dif^3\!r\;\frac{\mathrm{e}^{i\lih{k}\cdot\lih{r}}}{r^2}\left(\frac{\cos(2\mu r)}{\sinh(2\pi T r)} + \frac{|k_0|}{2\pi T}\right)
\mathrm{e}^{-|k_0|r}.
\end{equation}
From here we can proceed as in \cite{ArnoldZhai} by first subtracting the $T=0,\;\mu=0$ part (which contains the ultraviolet divergences) and then
evaluating the remaining integral. The final result is
\begin{eqnarray}
I_\mathrm{sunset} & = & -\frac{\mu^2}{64\pi^4}\left(\frac{1}{\epsilon} + 2\ln\frac{\mubar^2}{T^2}-4\ln 4\pi + 2\right) \nonumber \\
& & + \frac{i\mu T}{16\pi^3}\ln\frac{\Gamma\left(\frac{1}{2}-\frac{i\mu}{2\pi T}\right)}{\Gamma\left(\frac{1}{2}+\frac{i\mu}{2\pi T}\right)} + {\cal O}(\epsilon). \label{eq:pkpk}
\end{eqnarray}
We were unable to find whether this integral had been calculated at finite $\mu$ previously.

\subsection{Some properties of the required special functions}

The central function is the generalized zeta function $\zeta(z,q)$, defined in Eq. (\ref{zetadef}) for $\mathrm{Re}(z)>1$, 
and its derivative with respect to $z$, $\zeta'(z,q)$. From the integral representation
\begin{equation}
\zeta(z,q) = \frac{1}{\Gamma(z)}\int_0^\infty dt t^{z-1}\frac{\mathrm{e}^{(1-q)t}}{\mathrm{e}^t-1},
\end{equation}
valid at $\mathrm{Re}(z)>1,\; \mathrm{Re}(q)>0$, it is possible to analytically continue this function to the double complex plane $(z,q),\;z\neq 1$. For us it is sufficient to consider only the
half plane $\mathrm{Re}(q)>0$ where we can write
\begin{eqnarray}
\zeta(z,q) & = & \frac{1}{\Gamma(z)}\sum_{n=0}^\infty\frac{B_n(q)}{\Gamma(n+1)}\frac{(-1)^n}{z+n-1} \nonumber \\
& & + \frac{1}{\Gamma(z)}\int_1^\infty \mathrm{d}t t^{z-1}\frac{\mathrm{e}^{(1-q)t}}{\mathrm{e}^t-1}.
\end{eqnarray}
Here $B_n(q)$ are Bernoulli polynomials. This expression is well defined as long as $z\neq 1$ and $\mathrm{Re}(q)>0$ (the limit of $1/(\Gamma(z)(z+n-1))$ is well defined at $z=0,\;-1,\;-2,\dots$) 
and thus it can be used to define the analytic continuation of $\zeta(z,q)$ to that 
region. For the derivative we get
\begin{eqnarray}
\zeta'(z,q) & = & - \frac{1}{\Gamma(z)}\sum_{n=0}^\infty\frac{B_n(q)}{\Gamma(n+1)}\frac{(-1)^n}{(z+n-1)^2} \\
& & + \frac{1}{\Gamma(z)}\int_1^\infty dt t^{z-1}\ln t\frac{\mathrm{e}^{(1-q)t}}{\mathrm{e}^t-1} -\psi(z)\zeta(z,q). \nonumber
\end{eqnarray}
All the required special functions appearing in the integrals can be related to these functions. This is trivially true for the one-loop integrals by Eqs. (\ref{a1}). The logarithm of the gamma
functions present in the sunset integral Eq. (\ref{eq:pkpk}) can on the other hand be written as ($x = \mu/(\pi T)$)
\begin{equation}
\ln\frac{\Gamma\left(\frac{1}{2}-\frac{ix}{2}\right)}{\Gamma\left(\frac{1}{2}+\frac{ix}{2}\right)} = \zeta'\left(0,\frac{1}{2}-\frac{ix}{2}\right)-\zeta'\left(0,\frac{1}{2}+\frac{ix}{2}\right).
\label{eq:loggamma}
\end{equation}
Thus the mathematical properties of the non-polynomial part in $\mu/T$ of all the required integrals are given by the generalized zeta function.

The asymptotic behaviour of the integrals at the limits $\mu/T \rightarrow 0$ and $T/\mu\rightarrow 0$ is of interest. The small $\mu$ limit is a straightforward Taylor expansion and is not given
here explicitly. The large $\mu$ limit for the required functions is most easily obtained from Stirling's approximation for the gamma function:
\begin{equation}
\ln\frac{\Gamma\left(\frac{1}{2}-\frac{ix}{2}\right)}{\Gamma\left(\frac{1}{2}+\frac{ix}{2}\right)} = -\frac{ix}{2}\left(\ln \frac{x^2}{4}-2\right) -\frac{i}{6x} +{\cal O}\left(\frac{1}{x^3}\right).
\end{equation}
Thus
\begin{eqnarray}
2\mathrm{Re}\left[\psi\left(\frac{1}{2}-\frac{ix}{2}\right)\right] & = & 
2i\frac{d}{dx}\ln\frac{\Gamma\left(\frac{1}{2}-\frac{ix}{2}\right)}{\Gamma\left(\frac{1}{2}+\frac{ix}{2}\right)} \nonumber \\
& = & \ln \frac{x^2}{4} - \frac{1}{3x^2} + {\cal O}\left(\frac{1}{x^4}\right) \nonumber \\
2\mathrm{Re}\left[\zeta'\left(-1,\frac{1}{2}-\frac{ix}{2}\right)\right] & = & -\frac{x^2}{8}\left(\ln\frac{x^2}{4}-1\right) \nonumber \\
& & -\frac{1}{24}\ln x^2 +{\cal O}(x^0).
\end{eqnarray}
The last expansion can be derived with the help of Eq. (\ref{eq:loggamma}) after noting that the zeta function satisfies the relation
\begin{equation}
\frac{\partial}{\partial q} \zeta'(z,q) = -\zeta(z+1,q) -z\zeta'(z+1,q)
\end{equation}
and that $\zeta(0,q) = 1/2 - q$. Low temperature expansion of the integrals (\ref{eq:zetas}), (\ref{eq:psis}) and (\ref{eq:pkpk}) is now straightforward. We get
\begin{widetext}
\begin{eqnarray}
I^0_1  & = & -\frac{\mu^2}{8\pi^2}\left(1 + \frac{\pi^2}{3}\frac{T^2}{\mu^2} + \epsilon\left[\left(1+\frac{\pi^2}{3}\frac{T^2}{\mu^2}\right)\ln\frac{\mubar^2}{\mu^2}
+ 3-2\ln 2 + {\cal O}\left(\frac{T^2}{\mu^2}\right)\right]\right) + {\cal O}(\epsilon^2) \nonumber \\
I^0_2 & = & \frac{1}{16\pi^2}\left(\frac{1}{\epsilon} + \ln\frac{\mubar^2}{\mu^2} - 2\ln 2 + \frac{\pi^2}{3}\frac{T^2}{\mu^2} +{\cal O}\left(\frac{T^4}{\mu^4}\right)\right) 
+ {\cal O}(\epsilon) \nonumber \\
I_\mathrm{sunset} & = & -\frac{\mu^2}{64\pi^4}\left(\frac{1}{\epsilon} + \ln\frac{\mubar^2}{\mu^2} - 4\ln 2 + 6 -\frac{2\pi^2}{3}\frac{T^2}{\mu^2}
+ {\cal O}\left(\frac{T^4}{\mu^4}\right)\right) +{\cal O}(\epsilon).
\end{eqnarray}
\end{widetext}
Thus, although separate terms of the integrals diverge logarithmically at $T/\mu \rightarrow 0$, the integrals themselves are convergent in that limit, as they should.


\begin{thebibliography}{99}

%\cite{oneloop}
\bibitem{Kirzhnits}
D.~A.~Kirzhnits,
%``Weinberg Model In The Hot Universe,''
JETP Lett.\  {\bf 15} (1972) 529
[Pisma Zh.\ Eksp.\ Teor.\ Fiz.\  {\bf 15} (1972) 745].
D.~A.~Kirzhnits and A.~D.~Linde,
%``Macroscopic Consequences Of The Weinberg Model,''
Phys.\ Lett.\ B {\bf 42} (1972) 471.
%%CITATION = JTPLA,15,529;%%
%%CITATION = PHLTA,B42,471;%%

%\cite{Anderson}
\bibitem{Anderson}
G.~W.~Anderson and L.~J.~Hall,
%``The Electroweak phase transition and baryogenesis,''
Phys.\ Rev.\ D {\bf 45} (1992) 2685.
%%CITATION = PHRVA,D45,2685;%%

%\cite{Carrington}
\bibitem{Carrington}
M.~E.~Carrington,
%``The Effective potential at finite temperature in the Standard Model,''
Phys.\ Rev.\ D {\bf 45} (1992) 2933.
%%CITATION = PHRVA,D45,2933;%%

%\cite{Dine}
\bibitem{Dine}
M.~Dine, R.~G.~Leigh, P.~Y.~Huet, A.~D.~Linde and D.~A.~Linde,
%``Towards the theory of the electroweak phase transition,''
Phys.\ Rev.\ D {\bf 46} (1992) 550
[arXiv:hep-ph/9203203].
%%CITATION = HEP-PH 9203203;%%

%\cite{ArnoldEspinosa}
\bibitem{ArnoldEspinosa}
P.~Arnold and O.~Espinosa,
%``The Effective potential and first order phase transitions: Beyond leading-order,''
Phys.\ Rev.\ D {\bf 47} (1993) 3546
[Erratum-ibid.\ D {\bf 50} (1994) 6662]
[arXiv:hep-ph/9212235],
Z.~Fodor and A.~Hebecker,
%``Finite temperature effective potential to order g**4, lambda**2 and the electroweak phase transition,''
Nucl.\ Phys.\ B {\bf 432} (1994) 127
[arXiv:hep-ph/9403219].
%%CITATION = HEP-PH 9212235;%%
%%CITATION = HEP-PH 9403219;%%

%\cite{Generic}
\bibitem{Generic}
K.~Kajantie, M.~Laine, K.~Rummukainen and M.~E.~Shaposhnikov,
%``Generic rules for high temperature dimensional reduction and their application to the standard model,''
Nucl.\ Phys.\ B {\bf 458} (1996) 90
[arXiv:hep-ph/9508379].
%%CITATION = HEP-PH 9508379;%%

%\cite{ewphdg}
\bibitem{ewphdg}
K.~Kajantie, M.~Laine, K.~Rummukainen and M.~E.~Shaposhnikov,
%``The Electroweak Phase Transition: A Non-Perturbative Analysis,''
Nucl.\ Phys.\ B {\bf 466} (1996) 189
[arXiv:hep-lat/9510020],
%``Is there a hot electroweak phase transition at m(H) > approx. m(W)?,''
Phys.\ Rev.\ Lett.\  {\bf 77} (1996) 2887
[arXiv:hep-ph/9605288],
%``A non-perturbative analysis of the finite T phase transition in  SU(2) x U(1) electroweak theory,''
Nucl.\ Phys.\ B {\bf 493} (1997) 413
[arXiv:hep-lat/9612006],
F.~Karsch, T.~Neuhaus, A.~Patkos and J.~Rank,
%``Critical Higgs mass and temperature dependence of gauge boson masses in  the SU(2) gauge-Higgs model,''
Nucl.\ Phys.\ Proc.\ Suppl.\  {\bf 53} (1997) 623
[arXiv:hep-lat/9608087],
M.~Gurtler, E.~M.~Ilgenfritz and A.~Schiller,
%``Where the electroweak phase transition ends,''
Phys.\ Rev.\ D {\bf 56} (1997) 3888
[arXiv:hep-lat/9704013],
%%CITATION = HEP-LAT 9510020;%%
%%CITATION = HEP-PH 9605288;%%
%%CITATION = HEP-LAT 9612006;%%
%%CITATION = HEP-LAT 9608087;%%
%%CITATION = HEP-LAT 9704013;%%

%\cite{Csikor:1998eu}
\bibitem{Csikor:1998eu}
F.~Csikor, Z.~Fodor and J.~Heitger,
%``Endpoint of the hot electroweak phase transition,''
Phys.\ Rev.\ Lett.\  {\bf 82} (1999) 21
[arXiv:hep-ph/9809291].
%%CITATION = HEP-PH 9809291;%%

%\cite{univclass}
\bibitem{univclass}
K.~Rummukainen, M.~Tsypin, K.~Kajantie, M.~Laine and M.~E.~Shaposhnikov,
%``The universality class of the electroweak theory,''
Nucl.\ Phys.\ B {\bf 532} (1998) 283
[arXiv:hep-lat/9805013].
%%CITATION = HEP-LAT 9805013;%%

%\cite{magnetic}
\bibitem{magnetic}
K.~Kajantie, M.~Laine, J.~Peisa, K.~Rummukainen and M.~E.~Shaposhnikov,
%``The electroweak phase transition in a magnetic field,''
Nucl.\ Phys.\ B {\bf 544} (1999) 357
[arXiv:hep-lat/9809004].
%%CITATION = HEP-LAT 9809004;%%

%\cite{Orito}
\bibitem{Orito}
M.~Orito, T.~Kajino, G.~J.~Mathews and Y.~Wang,
%``Constraints on neutrino degeneracy from the cosmic microwave background  and primordial nucleosynthesis,''
Phys.\ Rev.\ D {\bf 65} (2002) 123504
[arXiv:astro-ph/0203352].
%%CITATION = ASTRO-PH 0203352;%%

%\cite{topdef}
\bibitem{topdef}
B.~Bajc, A.~Riotto and G.~Senjanovic,
%``Large lepton number of the universe and the fate of topological  defects,''
Phys.\ Rev.\ Lett.\  {\bf 81} (1998) 1355
[arXiv:hep-ph/9710415],
J.~McDonald,
%``Symmetry non-restoration via order 10**(-10) B and L asymmetries,''
Phys.\ Lett.\ B {\bf 463} (1999) 225
[arXiv:hep-ph/9907358],
B.~Bajc and G.~Senjanovic,
%``Large lepton number and high temperature symmetry breaking in MSSM,''
Phys.\ Lett.\ B {\bf 472} (2000) 373
[arXiv:hep-ph/9907552].
%%CITATION = HEP-PH 9710415;%%
%%CITATION = HEP-PH 9907358;%%
%%CITATION = HEP-PH 9907552;%%

%\cite{baulau}
\bibitem{baulau}
J.~McDonald,
%``Naturally large cosmological neutrino asymmetries in the MSSM,''
Phys.\ Rev.\ Lett.\  {\bf 84} (2000) 4798
[arXiv:hep-ph/9908300],
J.~March-Russell, H.~Murayama and A.~Riotto,
%``The small observed baryon asymmetry from a large lepton asymmetry,''
JHEP {\bf 9911} (1999) 015
[arXiv:hep-ph/9908396].
%%CITATION = HEP-PH 9908300;%%
%%CITATION = HEP-PH 9908396;%%

%\cite{Linde}
\bibitem{Linde}
A.~D.~Linde,
%``High Density And High Temperature Symmetry Behavior In Gauge Theories,''
Phys.\ Rev.\ D {\bf 14} (1976) 3345.
%%CITATION = PHRVA,D14,3345;%%

%\cite{Lindecond}
\bibitem{Lindecond}
A.~D.~Linde,
%``Classical Yang-Mills Solutions, Condensation Of W Mesons And Symmetry Of Composition Of Superdense Matter,''
Phys.\ Lett.\ B {\bf 86} (1979) 39.
%%CITATION = PHLTA,B86,39;%%

%\cite{BailinLove}
\bibitem{BailinLove}
D.~Bailin and A.~Love,
%``The Phase Transition In Electroweak Theory At Finite Density,''
Nucl.\ Phys.\ B {\bf 226} (1983) 493.
%%CITATION = NUPHA,B226,493;%%

%\cite{Ferrer}
\bibitem{Ferrer}
E.~J.~Ferrer, V.~de la Incera and A.~E.~Shabad,
%``Phase Transitions In The Electroweak Theory At Finite Temperature And Density,''
Phys.\ Lett.\ B {\bf 185} (1987) 407.
%%CITATION = PHLTA,B185,407;%%

%\cite{Kapusta}
\bibitem{Kapusta}
J.~I.~Kapusta,
%``Phase Diagram Of Electroweak Theory,''
Phys.\ Rev.\ D {\bf 42} (1990) 919.
%%CITATION = PHRVA,D42,919;%%

%\cite{Khlebnikov}
\bibitem{Khlebnikov}
S.~Y.~Khlebnikov and M.~E.~Shaposhnikov,
%``Melting of the Higgs vacuum: Conserved numbers at high temperature,''
Phys.\ Lett.\ B {\bf 387} (1996) 817
[arXiv:hep-ph/9607386].
%%CITATION = HEP-PH 9607386;%%

%\cite{Laine}
\bibitem{Laine}
M.~Laine and M.~E.~Shaposhnikov,
%``A remark on sphaleron erasure of baryon asymmetry,''
Phys.\ Rev.\ D {\bf 61} (2000) 117302
[arXiv:hep-ph/9911473].
%%CITATION = HEP-PH 9911473;%%

%\cite{Sannino:2002wp}
\bibitem{Sannino:2002wp}
F.~Sannino,
%``General structure of relativistic vector condensation,''
arXiv:hep-ph/0211367,
%%CITATION = HEP-PH 0211367;%%
%\cite{Sannino:2001fd}
%\bibitem{Sannino:2001fd}
F.~Sannino and W.~Schafer,
%``Relativistic massive vector condensation,''
Phys.\ Lett.\ B {\bf 527} (2002) 142
[arXiv:hep-ph/0111098].
%%CITATION = HEP-PH 0111098;%%

%\cite{RubaShapo}
\bibitem{RubaShapo}
V.~A.~Rubakov and M.~E.~Shaposhnikov,
%``Electroweak baryon number non-conservation in the early universe and in  high-energy collisions,''
Usp.\ Fiz.\ Nauk {\bf 166} (1996) 493
[Phys.\ Usp.\  {\bf 39} (1996) 461]
[arXiv:hep-ph/9603208].
%%CITATION = HEP-PH 9603208;%%
 
%\cite{Ginsparg}
\bibitem{Ginsparg}
P.~Ginsparg,
%``First Order And Second Order Phase Transitions In Gauge Theories At Finite Temperature,''
Nucl.\ Phys.\ B {\bf 170} (1980) 388,
%\cite{Appelquist:vg}
T.~Appelquist and R.~D.~Pisarski,
%``Hot Yang-Mills Theories And Three-Dimensional QCD,''
Phys.\ Rev.\ D {\bf 23} (1981) 2305.
%%CITATION = NUPHA,B170,388;%%
%%CITATION = PHRVA,D23,2305;%%
 
%\cite{Dunne}
\bibitem{Dunne}
G.~V.~Dunne,
%``Aspects of Chern-Simons theory,''
arXiv:hep-th/9902115.
%%CITATION = HEP-TH 9902115;%%

%\cite{Redlich}
\bibitem{Redlich}
A.~N.~Redlich and L.~C.~Wijewardhana,
%``Induced Chern-Simons Terms At High Temperatures And Finite Densities,''
Phys.\ Rev.\ Lett.\  {\bf 54} (1985) 970.
%%CITATION = PRLTA,54,970;%%

\bibitem{Rubakov} 
V.~A.~Rubakov and A.~N.~Tavkhelidze,
%``Stable Anomalous States Of Superdense Matter In Gauge Theories,''
Phys.\ Lett.\ B {\bf 165} (1985) 109,
V.~A.~Rubakov,
%``On The Electroweak Theory At High Fermion Density,''
Prog.\ Theor.\ Phys.\  {\bf 75} (1986) 366,
D.~V.~Deryagin, D.~Y.~Grigoriev and V.~A.~Rubakov,
%``Inhomogeneous W Boson Condensates In The Standard Electroweak Theory At High Fermionic Densities,''
Phys.\ Lett.\ B {\bf 178} (1986) 385.
%%CITATION = PHLTA,B165,109;%%
%%CITATION = PTPKA,75,366;%%
%%CITATION = PHLTA,B178,385;%%

%\cite{aleksi}
\bibitem{aleksi}
A.~Vuorinen,
%``Quark number susceptibilities of hot QCD up to g**6 ln(g),''
Phys.\ Rev.\ D {\bf 67} (2003) 074032
[arXiv:hep-ph/0212283].
%%CITATION = HEP-PH 0212283;%%

%\cite{ArnoldZhai}
\bibitem{ArnoldZhai}
P.~Arnold and C.~X.~Zhai,
%``The Three Loop Free Energy For Pure Gauge QCD,''
Phys.\ Rev.\ D {\bf 50} (1994) 7603
[arXiv:hep-ph/9408276].
%``The Three loop free energy for high temperature QED and QCD with fermions,''
Phys.\ Rev.\ D {\bf 51} (1995) 1906
[arXiv:hep-ph/9410360].
%%CITATION = HEP-PH 9408276;%%
%%CITATION = HEP-PH 9410360;%%

\end{thebibliography}
\end{document}